\documentclass[aps,prd,preprintnumbers,showpacs,showkeys,nofootinbib,
superscriptaddress,fleqn,floatfix,tightenlines,10pt]{revtex4-1}
\usepackage[makeroom]{cancel}
\usepackage{amsmath,amsfonts,amssymb,amscd,amsxtra,amsthm}
\usepackage{graphicx}   
\usepackage{epstopdf}
\usepackage{dcolumn}    
\usepackage{bm}         
\usepackage{slashed}
\usepackage{cancel}
\usepackage{float}
\usepackage{mathtools,amsbsy,amstext}
\usepackage[toc,page]{appendix}

\usepackage[utf8]{inputenc}
\usepackage{booktabs}
\usepackage[normalem]{ulem} 
\usepackage[dvipsnames]{xcolor} 
\usepackage{tabularx}
\usepackage{enumitem}
\usepackage{array}
\usepackage{tikz}
\usepackage{multirow}
\renewcommand\sout{\bgroup \color{red} \ULdepth=-.5ex \ULset}

\begin{document}
\preprint{PKNU-NuHaTh-2020-04}
\title{Photoproduction of $\Lambda^*$ and $\Sigma^*$ resonances with $J^P=1/2^-$
off the proton}
\author{Sang-Ho Kim}
\email[E-mail: ]{shkim@pknu.ac.kr}
\affiliation{Department of Physics, Pukyong National University (PKNU),
Busan 48513, Korea}

\author{K.\,P.\,Khemchandani}
\email[E-mail: ]{kanchan.khemchandani@unifesp.br}
\affiliation{Universidade Federal de S\~ao Paulo, C.P. 01302-907, S\~ao Paulo,
Brazil.}

\author{A.\,Mart\'inez Torres}
\email[E-mail: ]{amartine@if.usp.br}
\affiliation{Universidade de Sao Paulo, Instituto de Fisica, C.P. 05389-970,
Sao Paulo, Brazil.}

\author{Seung-il Nam}
\email[E-mail: ]{sinam@pknu.ac.kr}
\affiliation{Department of Physics, Pukyong National University (PKNU), Busan
48513, Korea}
\affiliation{Asia Pacific Center for Theoretical Physics (APCTP), Pohang 37673,
Korea}

\author{Atsushi Hosaka}
\email[E-mail: ]{hosaka@rcnp.osaka-u.ac.jp}
\affiliation{Research Center for Nuclear Physics (RCNP), Osaka University,
Ibaraki, Osaka, 567-0047, Japan}
\affiliation{Advanced Science Research Center, Japan Atomic Energy Agency
(JAEA), Tokai 319-1195, Japan}
\date{\today}
\begin{abstract}
We study the photoproduction of the $\Lambda(1405)$ and $\Sigma(1400)$ hyperon resonances, the latter of which is not a well established state. We evaluate the $s$-, $t$- and $u$-channel diagrams in the Born approximation by employing the effective Lagrangians. A new ingredient is the inclusion of a nucleon resonance $N^*(1895)$ that is dynamically generated with predictions for its coupling to the $K\Lambda(1405)$ and $K\Sigma(1400)$ channels. To extend the applicability of the model to energies beyond the threshold region, we consider a Regge model for the  $t$-channel $K$- and $K^*$-exchanges. Our results are in good agreement with the CLAS data available on $\Lambda(1405)$, while for $\Sigma(1400)$ we predict observables for its production. We also provide polarization observables for both  hyperon productions, which can be useful in future experimental investigations. The present study provides  new information on the nucleon resonance $N^*(1895)$  which can be an alternative source for generating the hyperon resonances $\Lambda(1405)$ and $\Sigma(1400)$.
\end{abstract}
\maketitle

\section{Introduction}
Understanding the properties of low-lying hyperon resonances is of special interest to the hadron physics community, since there are indications that their structure is much richer than the three valence quark composition. For instance, the mass of the first excited state with isospin 0, $\Lambda(1405)$, is lower than its non-strange counterpart, $N^*(1535)$. The properties of $\Lambda(1405)$ have been studied  in various models, and it is suggested that $\Lambda(1405)$ is a state arising from hadron dynamics and can be interpreted as a hadronic molecular state (for a recent review see Ref.~\cite{Mai:2020ltx} or Refs.~\cite{Jido:2003cb,Hyodo:2011ur,GarciaRecio:2002td,Dalitz:1967fp,Magas:2005vu,Mai:2012dt,Kaiser:1995eg,osetramos,Oller:2000fj,MuellerGroeling:1990cw} for some top-cited works). Different studies attribute a double pole structure in the complex energy plane to the $\Lambda(1405)$ state seen on the real axis. However, such a double pole nature is still under discussions~\cite{Klempt:2020lkh}.

An interesting suggestion has been also made for the isovector sector, while the arguments are not yet fully settled.  Some studies find evidence for the existence of a $J^P=1/2^-$ $\Sigma$ state in the mass region of $\Lambda(1405)$ while others do not~\cite{Oller:2000fj,Guo,Wu:2009tu,Wu:2009nw,Gao:2010hy,Xie:2014zga,Xie:2017xwx,Moriya:2013eb,Roca:2013cca,Klempt:2020lkh}. In a recent article~\cite{Khemchandani:2018amu}, some of the coauthors of the present work studied the strangeness $-1$ coupled channel interactions by constraining the model parameters through a $\chi^2$-fit to the relevant experimental data. Besides constraining the parameters, more diagrams were considered in Ref.~\cite{Khemchandani:2018amu}  as compared to the former study~\cite{Khemchandani:2011mf}. In both works, a particular feature is that the pseudoscalar and vector mesons are considered in the coupled channels space. Although the vector-baryon thresholds lie away from the mass region of $\Lambda(1405)$ and, thus, do not play a crucial role in its generation, important information is obtained within such a formalism, i.e., the couplings of the vector-baryon channels to $\Lambda(1405)$ (besides the generation of other hyperon states). Such information is valuable for the study of other processes like the photoproduction of hyperon resonances. In Ref.~\cite{Khemchandani:2018amu}, two poles were found in relation with $\Lambda(1405)$, in agreement with  the analysis~\cite{Roca:2013cca,Mai:2014xna} of the data on the  electroproduction and photoproduction of $\Lambda(1405)$~\cite{Moriya:2013eb,Lu:2013nza,Moriya:2013hwg}. Additionally, a $1/2^-$ $\Sigma$ state, with mass around 1400 MeV, in the isovector sector was also found. We shall refer to this state as $\Sigma(1400)$  in the following discussions.

With the idea of bringing more  information on the topic, in the present work, we study the photoproduction of light hyperons off the proton, i.e., $\gamma p \to K^+ \Sigma^0(1400)$ and $\gamma p \to K^+ \Lambda(1405)$. Indeed the data on the photoproduction of $\Lambda(1405)$ are already available from the CLAS Collaboration~\cite{Moriya:2013eb,Moriya:2013hwg}, and the data with better statistics are expected to be released from the experiments planned at the ELSA facility in Bonn~\cite{Scheluchin:2020mhn}. The aim of the latter facility is to better establish the nature of $\Lambda(1405)$. We find that the results 
obtained in Ref.~\cite{Khemchandani:2018amu} are useful in reproducing the $K^+ \Lambda(1405)$ production data from the CLAS/JLab~\cite{Moriya:2013hwg}. Further, to motivate similar experimental studies of $\Sigma(1400)$, we predict the cross sections for its production in photon-proton collisions. We also provide the results on the asymmetries and polarized cross sections for $K^+\Lambda(1405)$ as well as $K^+\Sigma(1400)$ productions, which can be useful for future experimental investigations. 



To accomplish the above mentioned goals, we consider the $s$-, $t$- and $u$-channel Born diagrams and employ an effective Lagrangian approach where the couplings of the different vertices are obtained mainly from Ref.~\cite{Khemchandani:2018amu}. Besides considering the nucleon exchange in the $s$-channel, we also include $N^*(1895)$, which lies extremely close to the $KY^*$ thresholds, where $Y^*$ denotes $\Lambda(1405)$ or $\Sigma(1400)$. We do not require to introduce any unknown parameters when considering the $N^*(1895)$ exchange, since its decay to the $KY^*$ channel was recently studied in Ref.~\cite{Khemchandani:2020exc}. Furthermore, we determine the electromagnetic couplings of $N^*(1895)$ (as well as of $\Lambda(1405)$ and $\Sigma(1400)$) by using the vector dominance model, where the required vector-baryon couplings to the resonant states are taken from Refs.~\cite{Khemchandani:2018amu,Khemchandani:2013nma}. We find that the $N^*(1895)$ exchange plays an important role in describing the $\Lambda(1405)$ photoproduction data near the threshold region. To describe the cross sections for energies away from the threshold region of the reactions, the effective Lagrangian approach is complemented with  a Regge model~\cite{Kim:2017nxg} in which the $t$-channel $K$- and $K^*$-Reggeon exchange processes are considered.

The paper is organized as follows. In Sec.~\ref{SecII}, we describe our theoretical framework. In Sec.~\ref{SecIII}, we show and discuss our numerical results of the total and differential cross sections for $\gamma p \to K^+ \Lambda(1405)$ and $\gamma p \to K^+ \Sigma^0(1400)$. We also predict some asymmetries and polarized cross sections. The final section is devoted to the summary.

\section{Theoretical Framework}
\label{SecII}
We start our discussions by introducing the effective Lagrangians for the photoproduction of the hyperon resonances $\Lambda^*\equiv \Lambda(1405)$ and $\Sigma^*\equiv \Sigma(1400)$, together with the basics of the Regge model for the $t$-channel exchange processes. Next, we provide the formalism to evaluate the strong vertices $N^*(1895) \to K \Lambda(1405)$, $K \Sigma(1400)$, and the radiative decays of $N^*(1895)$, $\Lambda(1405)$, and $\Sigma(1400)$, which are necessary to determine the $\gamma p\to N^*(1895)\to KY^*$ cross sections.

\subsection{Effective Lagrangians and Regge model}
\begin{figure}[htp]
\centering
\includegraphics[width=9.5cm]{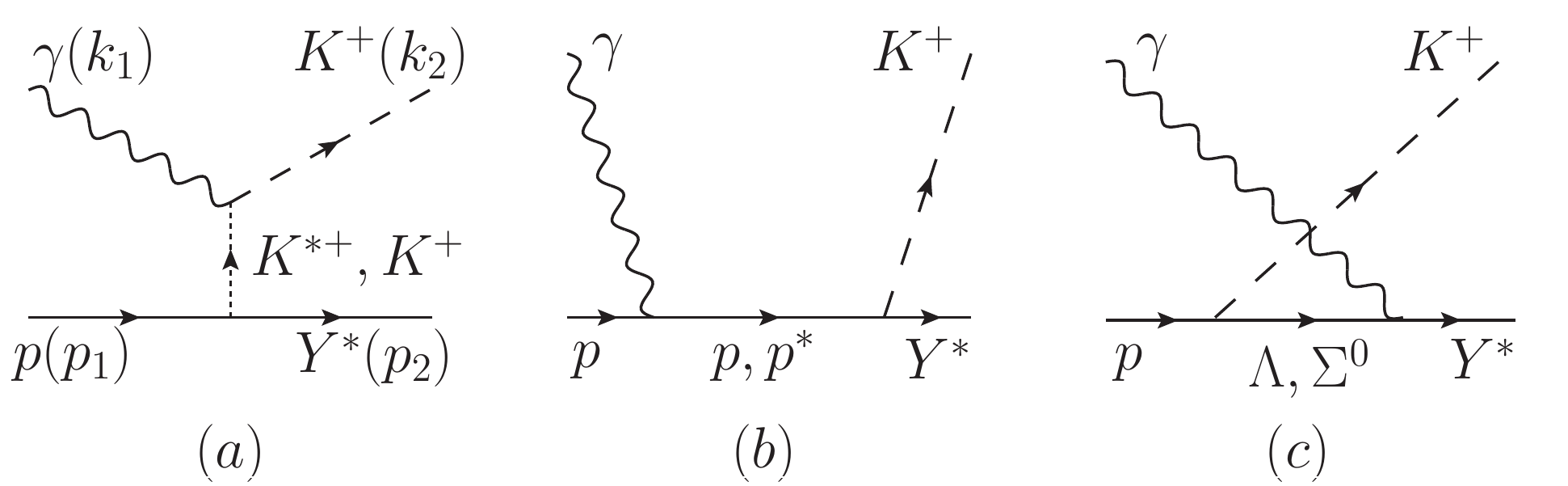}
\caption{(a) $t$- (b) $s$- and (c) $u$-channel Feynman diagrams for $\gamma p
\to K^+ Y^*$, where $Y^* = \Lambda^*$ or $\Sigma^*$ and $k_1$ ($k_2$), $p_1$ ($p_2$) are the four-momenta assigned to the particles in the initial (final) state.}
\label{fig1}
\end{figure}

Within our approach, the production mechanism of the $Y^*$ resonances in the reaction $\gamma p \to K^+ Y^*$ consists of the standard
$t$-, $s$-, and $u$-channel Born terms combined with the $s$-channel resonance exchange  
as shown in Fig.~\ref{fig1}.

The effective Lagrangians for the electromagnetic (EM) interaction vertices shown in Fig.~\ref{fig1} are
written as
\begin{align}
\mathcal L_{\gamma K K} &=
-ie [  K^\dagger (\partial_\mu K) - (\partial_\mu K^\dagger) K ] A^\mu ,       \cr
\mathcal L_{\gamma K K^*} &=
g_{\gamma K K^*}^c \epsilon^{\mu\nu\alpha\beta} \partial_\mu A_\nu
[ (\partial_\alpha K_\beta^{*-}) K^+ + K^- (\partial_\alpha K_\beta^{*+} )] ,    \cr
\mathcal L_{\gamma NN} &=
- e \bar N \left[ \gamma_\mu \frac{1+\tau_3}{2} - \frac{\kappa_N}{2M_N}
\sigma_{\mu\nu}\partial^\nu \right] A^\mu N ,                                \cr
\mathcal L_{\gamma Y Y^*} &=
\frac{e\mu_{Y^*Y}}{2M_N} \bar Y \gamma_5 \sigma_{\mu\nu} \partial^\nu
A^\mu Y^* + \mathrm{H.c.},
\label{eq:Lag:EM}
\end{align}
where $A^\mu$ is the photon field, $e$ is the unit electric charge and $Y$ denotes the field for the ground-state $\Lambda (1116)$ or $\Sigma^0(1192)$. In Eq.~(\ref{eq:Lag:EM}),
the coupling constant $g_{\gamma K K^*}^c$ is determined from the experimental data
for $\Gamma_{K^{*+} \to K^+ \gamma}$~\cite{Zyla:2020zbs}, which gives a value of 0.254
$\mathrm{GeV}^{-1}$, and $\kappa_p$ = 1.79~\cite{Zyla:2020zbs} is the proton anomalous magnetic moment. As for the transition magnetic moments $\mu_{Y^*Y}$, we refer the reader to the next subsection for the details on its determination.

For the strong interaction vertices shown in Fig.~\ref{fig1}, the corresponding effective Lagrangians read as
\begin{align}
\mathcal L_{K N Y} &=
- i g_{K N Y} \bar N \gamma_5 Y K + \mathrm{H.c.},                           \cr
\mathcal L_{K N Y^*} &=
g_{K N Y^*} \bar K \bar Y^* N + \mathrm{H.c.},                                \cr
\mathcal L_{K^* N Y^*} &=
i\frac{g_{K^* N Y^*}}{\sqrt3} \bar K^{*\mu} \bar Y^* \gamma_\mu \gamma_5 N
+ \mathrm{H.c.},
\label{eq:Lag:Strong}
\end{align}
where the strong coupling $g_{K N(\Lambda,\,\Sigma^0)}$ is given by $(-13.4,\,4.09)$
from the Nijmegen soft-core potential (NSC97a)~\cite{Stoks:1999bz,Rijken:1998yy}.
 We take the values for the $Y^*\bar K N$ and $Y^*\bar K^*N$ couplings from Ref.~\cite{Khemchandani:2018amu}, where these former couplings are determined by solving coupled channel scattering equations for the pseudoscalar- and vector-baryon channels. The factor $1/\sqrt{3}$ in $\mathcal{L}_{K^*NY^*}$ takes into account
the fact that the Lagrangian has a spin structure while
  the couplings in
 Refs.~\cite{Khemchandani:2018amu,Khemchandani:2013nma} are obtained by evaluating  the residue of the $t$-matrices projected on spin 1/2. We list the values of the $Y^*\bar K N$ and
$Y^*\bar K^*N$ couplings in  Table~\ref{TAB1}.
\begin{table}[htp]
\begin{tabular}{c|c|c|c|c}
\hline\hline
&\multicolumn{2}{c|}{$\Lambda$(1405)}
& \multicolumn{2}{c}{$\Sigma$'s around 1400 MeV} \\
&$1385^{\pm 5} - i\,124^{\pm 10}$&$1426^{\pm 1} -i\,15^{\pm 2}$&
$-$&$1399^{\pm 35} -i\,36^{\pm 9}$   \\   \hline
$\bar K N$&$0.66^{\pm 0.35} - i\,1.93^{\pm 0.12}$&
$2.43^{\pm 0.16} + i\,0.63^{\pm 0.23}$&
$\hspace{5em}-\hspace{5em}$
&$\hspace{0.8em}0.50^{\pm 0.29} + i\,0.33^{\pm 0.18}$   \\
$\bar K^* N$&$0.62^{\pm 0.28} - i\,0.18^{\pm 0.14}$&
$0.04^{\pm 0.36} + i0.23\,^{\pm 0.19}$&
$-$&$-3.46^{\pm 0.21} - i\,0.06^{\pm 0.15}$   \\
\hline \hline
\end{tabular}
\caption{Pole positions of $\Lambda(1405)$ and $\Sigma(1400)$ and their couplings
to $\bar K N$ and $\bar K^* N$ (in the isospin base)~\cite{Khemchandani:2018amu}.
Note that the $K^{(*)} N \Lambda^*$ and $K^{(*)+} \bar p \Lambda^*$ couplings
are related by the Clebsh-Gordan coefficient $\frac{1}{\sqrt2}$, following the
convention $|K^-\rangle=-|I=1/2, I_z=-1/2\rangle$.
Also the $K^{(*)} N \Sigma^*$ and $K^{(*)+} \bar p \Sigma^{*0}$ couplings are related
by a factor $-\frac{1}{\sqrt2}$.}
\label{TAB1}
\end{table}
A comment on the $\bar K^*NY^*$ couplings given in Table~\ref{TAB1} is here in order. The values used in this work, as taken  from Ref.~\cite{Khemchandani:2018amu}, are different to those obtained in the former work~\cite{Khemchandani:2011mf}. This is because the formalism in Ref.~\cite{Khemchandani:2018amu} was updated by including  $s$- and $u$-channel diagrams for the pseudoscalar-baryon interactions, which were not taken into account in Ref.~\cite{Khemchandani:2011mf}. Additionally, a fit to experimental data was made in Ref.~\cite{Khemchandani:2018amu} to constrain the model parameters. The data considered were cross sections of the processes: $K^- p \to K^- p$, $K^- p \to \bar K^0 n$, $K^- p \to \eta \Lambda$, $K^- p \to \pi^0 \Lambda$, $K^- p \to \pi^0 \Sigma^0$, $K^- p \to \pi^\pm \Sigma^\mp$.  Data on the energy level shift and width of the $1s$ state of the kaonic hydrogen were also considered in the fit in Ref.~\cite{Khemchandani:2018amu}. However, it must be mentioned that data on final states consisting of vector mesons were not taken into account. Such considerations may update the values given in Table~\ref{TAB1}, in future.

In the Particle Data Group (PDG) 2020 edition~\cite{Zyla:2020zbs}, a new resonance
$\Lambda(1380,1/2^-)$ is included while the features of $\Lambda(1405,1/2^-)$ are
kept to be almost the same.
Thus, in this work, the information of the $g_{KN\Lambda^*}$ and $g_{K^*N\Lambda^*}$
are taken from the  higher pole given in Table~\ref{TAB1}.
It should be also mentioned that we use the average values of the couplings listed in Table~\ref{TAB1}.

For the exchange of a ground state hadron $h$, the individual amplitudes can be
written in the form
\begin{align}
M_h = \bar u_{Y^*} \mathcal M_h^\mu \epsilon_\mu u_N,\label{eq:BornAmp1}
\end{align}
where
\begin{align}
\mathcal M_K^\mu &= -2 \frac{eg_{KNY^*}}{t-M_K^2} k_2^\mu,   \cr
\mathcal M_N^\mu &= -\frac{eg_{KNY^*}}{s-M_N^2} (\rlap{/}{q_s}+M_N)
\left [ \gamma^\mu + \frac{i\kappa_p}{2M_N} \sigma^{\mu\nu} k_{1\nu} \right ],  \cr
\mathcal M_{K^*}^\mu &= -\frac{i}{\sqrt3}
\frac{g_{\gamma KK^*} g_{K^*NY^*}}{t-M_{K^*}^2}
\epsilon^{\mu\nu\alpha\beta}  \gamma_\nu \gamma_5 k_{1\alpha} k_{2\beta} ,          \cr
\mathcal M_Y^\mu &= \frac{e\mu_{Y^*Y}}{2M_N} \frac{g_{KNY}}{u-M_Y^2}
\sigma^{\mu\nu} k_{1\nu} (\rlap{/}{q_u}-M_Y),
\label{eq:BornAmp}
\end{align}
and $q_{s,u}$ denote the four momenta of the exchanged particles, i.e.,
$q_s=k_1+p_1$ and $q_u=p_2-k_1$.  In Eq.~(\ref{eq:BornAmp1})
$u_p$ and $u_{Y^*}$ stand for the Dirac spinors of the incoming proton and
 outgoing $Y^*$ hyperon, respectively, and $\epsilon_\mu$ is the polarization
vector for the incident photon.

We employ a hybridized Regge model for the dominant $t$-channel $K$- and
$K^*$-exchanges in order to extend the applicable energy region.
To do this, the Feynman propagators are replaced by the Regge
ones~\cite{Guidal:1997hy}
\begin{align}
P_K^{\mathrm{Feyn}} = \frac{1}{t-M_K^2} & \to P_K^{\mathrm{Regge}}
(s,t) = \left( \frac{s}{s_0^K} \right)^{\alpha_K(t)}
\frac{\pi\alpha_K'}{\sin[\pi\alpha_K(t)]}
\left\{ \begin{array}{c} 1 \\ e^{-i\pi\alpha_K(t)} \end{array} \right\}
\frac{1}{\Gamma[1+\alpha_K(t)]},                                          \cr
P_{K^*}^{\mathrm{Feyn}} = \frac{1}{t-M_{K^*}^2} & \to P_{K^*}^{\mathrm{Regge}}
(s,t) = \left( \frac{s}{s_0^{K^*}} \right)^{\alpha_{K^*}(t)-1}
\frac{\pi\alpha'_{K^*}}{\sin[\pi\alpha_{K^*}(t)]}
\left\{ \begin{array}{c} 1 \\ e^{-i\pi\alpha_{K^*}(t)} \end{array} \right\}
\frac{1}{\Gamma[\alpha_{K^*}(t)]},
\label{eq:ReggeProp}
\end{align}
such that we can explore higher photon-energy regions.
The Regge trajectories read~\cite{Guidal:1997hy}
\begin{align}
\alpha_K(t)= 0.7\, \mathrm{GeV^{-2}} (t - M_K^2), \hspace{1em}
\alpha_{K^*}(t)= 1 + 0.83\, \mathrm{GeV^{-2}} (t - M_{K^*}^2),
\label{eq:ReggeTraj}
\end{align}
where $\alpha'_{K,K^*} \equiv \partial \alpha_{K,K^*}(t)/\partial t$.
The energy scale parameters in Eq.~(\ref{eq:ReggeProp}) are determined to be
$s_0^K = 3.0\, \mathrm{GeV^2}$ and  $s_0^{K^*}= 1.5 \, \mathrm{GeV^2}$.
Since the $K$ and $K^*$ Regge trajectories are known to be degenerate, we
consider both a constant $(1)$ and a rotating $(e^{-i\pi\alpha_{K,K^*}(t)})$ phase.

In the effective Lagrangian framework  the individual
amplitudes for the $K$ exchange and the electric part of the $N$ exchange do not satisfy the
gauge invariance, but their sum does when the scalar (S) meson-baryon
coupling scheme is used as given in Eq.~(\ref{eq:Lag:Strong}) for
$\mathcal L_{KNY^*}$.
Thus we use the following prescription for the full amplitude:
\begin{align}
\mathcal M  =
\left( \mathcal M_K + \mathcal M_N^{\mathrm{elec}} \right) (t - M_K^2)
P_K^{\mathrm{Regge}} + \mathcal M_{K^*} (t - M_{K^*}^2) P_{K^*}^{\mathrm{Regge}} +
\mathcal M_N^{\mathrm{magn}} F_N +
\sum_{Y=\Lambda,\Sigma^0} \mathcal M_Y F_Y,
\label{eq:Amp:SRegge}
\end{align}
where the magnetic part of the $N$ exchange is self gauge invariant.
Here, we introduce a form factor for the baryon exchange process
\begin{align}
F_B(q^2) =\left[ \frac{\Lambda_B^4}{\Lambda_B^4 + \left(q^2-M_B^2\right)^2}
\right]^2,
\label{eq:FF}
\end{align}
where $q_{s,u}^2 =(s,u)$ denote the Mandelstam variables. 
In case of the diagram involving the $N^*$-exchange, we introduce the Gaussian
form factor
\begin{align}
F_{N^*}(s) = \mathrm{exp}\left[ -\frac{(s-M_{N^*}^2)^2}{\Lambda_{N^*}^4} \right]
\label{eq:GauFF}
\end{align}
such that the corresponding amplitude decreases more sharply with energy than it
would with the form factor of Eq.~(\ref{eq:FF}).
The latter choice is based on the consideration that the excited baryons, like
$N^*(1895)$, which can be understood as moleculelike states, have more spatially
extended structure than the ground state hadrons.

Note that when we use the vector (V) meson-baryon coupling scheme as
$\mathcal L_{KNY^*}^V = if_{K N Y^*} \partial^\mu \bar K \bar Y^* \gamma_\mu N$ with
the relation $f_{K N Y^*}=g_{K N Y^*}/(M_{Y^*}-M_N)$, a contact term is additionally
needed to satisfy the gauge-invariance condition and the first term of
Eq.~(\ref{eq:Amp:SRegge}) turns out to be equivalent to the following amplitude:
\begin{align}
\mathcal M^V =
\left( \mathcal M_K^V + \mathcal M_N^{\mathrm{elec,V}} + \mathcal M_C \right)
(t-M_K^2) P_K^{\mathrm{Regge}}.
\label{eq:PVRegge}
\end{align}
The contact term is obtained from the Lagrangian $\mathcal L_{KNY^*}^V$ by the
substitution $\partial^\mu \to -ieA^\mu$.

\subsection{Strong decay: $N^*(1895) \to K\Lambda(1405),\,
 K\Sigma(1400)$}
In addition to the Born-term contribution mentioned in the previous subsection, it is important to consider the exchange of a nucleon resonance in the $s$-channel as shown in Fig.~\ref{fig1}(b). The relevance of such diagrams was pointed out in Ref.~\cite{Kim:2017nxg}. In this work,  we consider the exchange of $N^*(1895)$.  The motivation for such a consideration is twofold: (1) the state $N^*(1895)$ lies very close to the $K\Lambda(1405)$ and $K\Sigma(1400)$ thresholds and (2) the partial decay widths of $N^*(1895)$ to $K\Lambda(1405)$ and $K\Sigma(1400)$ have been recently  determined in Ref.~\cite{Khemchandani:2020exc}. The results in Ref.~\cite{Khemchandani:2020exc} can be used to calculate the coupling constants which can be implemented in the present study of the photoproduction of $\Lambda(1405)$ and $\Sigma(1400)$.

To determine the contribution of $N^*(1895)$, which has spin and parity $J^P = 1/2^-$, we write the following effective Lagrangian for the strong interaction
vertex
\begin{align}
\mathcal L_{KY^*N^*}^{1/2^-} = -i g_{KY^*N^*} \bar K \bar Y^* \gamma_5 N^* +
\mathrm{H.c.},
\label{eq:NsSt}
\end{align}
where the coupling constant $g_{KY^*N^*}$ is determined such that  the $N^*\to K Y^*$ decay widths obtained in Ref.~\cite{Khemchandani:2020exc} can be reproduced through 
\begin{align}\nonumber
\Gamma_{N^* \to K Y^*} = \frac{1}{N}\int
\limits_{\left(M_{N^*}-2\Gamma_{N^*}\right)^2}^{\left(M_{N^*}+2\Gamma_{N^*}\right)^2}
d\tilde m^2&\left(-\frac{1}{\pi}\right) \text{Im}
\left[ \frac{1}{\tilde m^2-M_{N^*}^2+i M_{N^*} \Gamma_{N^*}} \right]
\frac{|\,\vec p_\text{c.m.}\,(\tilde m)|}{4 \pi \tilde m} |g_{K Y^*N^*}|^2
[ E_{Y^*}(\tilde m)- M_{Y^*} ] \\
& \times \Theta\left(\tilde m-M_K-M_{Y^*}\right). \label{conv}
\end{align}
Here,
$|\,\vec p_\text{c.m.}\,(\tilde m)|$ and $E_{Y^*}(\tilde m)$ denote the modulus of the center-of mass (c.m.) momentum and the energy of the hyperon $Y^*$ calculated as a function of the integration variable, respectively. $N$ is a normalization factor
\begin{align}
N = \int\limits_{\left(M_{N^*}-2\Gamma_{N^*}\right)^2}^{\left(M_{N^*}+2\Gamma_{N^*}\right)^2} d\tilde m^2\left(-\frac{1}{\pi}\right) \text{Im}\left[\frac{1}{\tilde m^2-M_{N^*}^2+i M_{N^*} \Gamma_{N^*}}\right].
\end{align}
The meaning of the integration in Eq.~(\ref{conv}) is to calculate the partial decay width of $N^*\to KY^*$ by taking into account the finite width of $N^*$ itself~\cite{Khemchandani:2020exc}.  It should be recalled that $N^*(1895)$ is associated with two poles in the complex energy plane in Ref.~\cite{Khemchandani:2020exc}, based on the findings of Ref.~\cite{Khemchandani:2013nma}. The corresponding poles positions  are $M_{N^*}-i\Gamma_{N^*}/2 = 1801-i96$~MeV and $1912-i54$~MeV. We would like to emphasize that although the real part of the former pole lies below the $KY^*$ threshold, its exchange can still be considered due to its width, which makes that the tail of the resonance allows for its decay to $KY^*$. Indeed, the influence of an $N^*$ on the photoproduction data was discussed in Ref.~\cite{MartinezTorres:2009cw}, where the nominal mass of the $N^*$ lies below the reaction threshold.
Using Eq.~(\ref{conv}), we  determine the absolute values of $g_{KY^*N^*}$ which reproduce the partial widths obtained in Ref.~\cite{Khemchandani:2020exc}.
The values  of  the couplings to be used in Eq.~(\ref{eq:NsSt}) are given in Table~\ref{TAB2}, where
\begin{table}[htp]
\begin{tabular}{ccc}
\hline\hline
Decay channel & Partial width [MeV] & $|g_{KY^*N^*}|$   \\
\hline
$N_1^{*} \to K \Lambda_1^*$ & 10.4 $\pm$ 1.3 & 10.9   \\
$N_1^{*} \to K \Lambda_2^*$ & 6.4 $\pm$ 0.8 & 11.1   \\
$N_1^{*} \to K \Sigma^{*}$ & 11.4 $\pm$ 1.5 & 12.5   \\
$N_2^{*} \to K \Lambda_1^*$ & 1.9 $\pm$ 0.1 & 3.6   \\
$N_2^{*} \to K \Lambda_2^*$ & 1.1 $\pm$ 0.2 & 4.0   \\
$N_2^{*} \to K \Sigma^{*}$ & 12.1 $\pm$ 1.2 & 10.2   \\
\hline\hline
\end{tabular}
\caption{Partial decay widths~\cite{Khemchandani:2020exc} and the extracted
couplings $g_{KY^*N(1895)}$. Here, $N^*$, $\Lambda^*$ and $\Sigma^*$ denote the states $N^*(1895)$, $\Lambda(1405)$ and $\Sigma(1400)$, respectively. The subscripts refer to the two pole nature of  $N^*(1895)$ and $\Lambda(1405)$.}
\label{TAB2}
\end{table}
the labels $N^*_1$ and $N^*_2$ refer to the two  poles associated with $N^*(1895)$~\cite{Khemchandani:2020exc,Khemchandani:2013nma}. Similarly, the subscripts on $\Lambda^*$ indicate the two poles related to $\Lambda(1405)$~\cite{Khemchandani:2018amu}. The coupling constants given in Table~\ref{TAB2} correspond to the central values of the widths listed alongside.

\begin{figure}[htp]
\centering
\includegraphics[width=5.45cm]{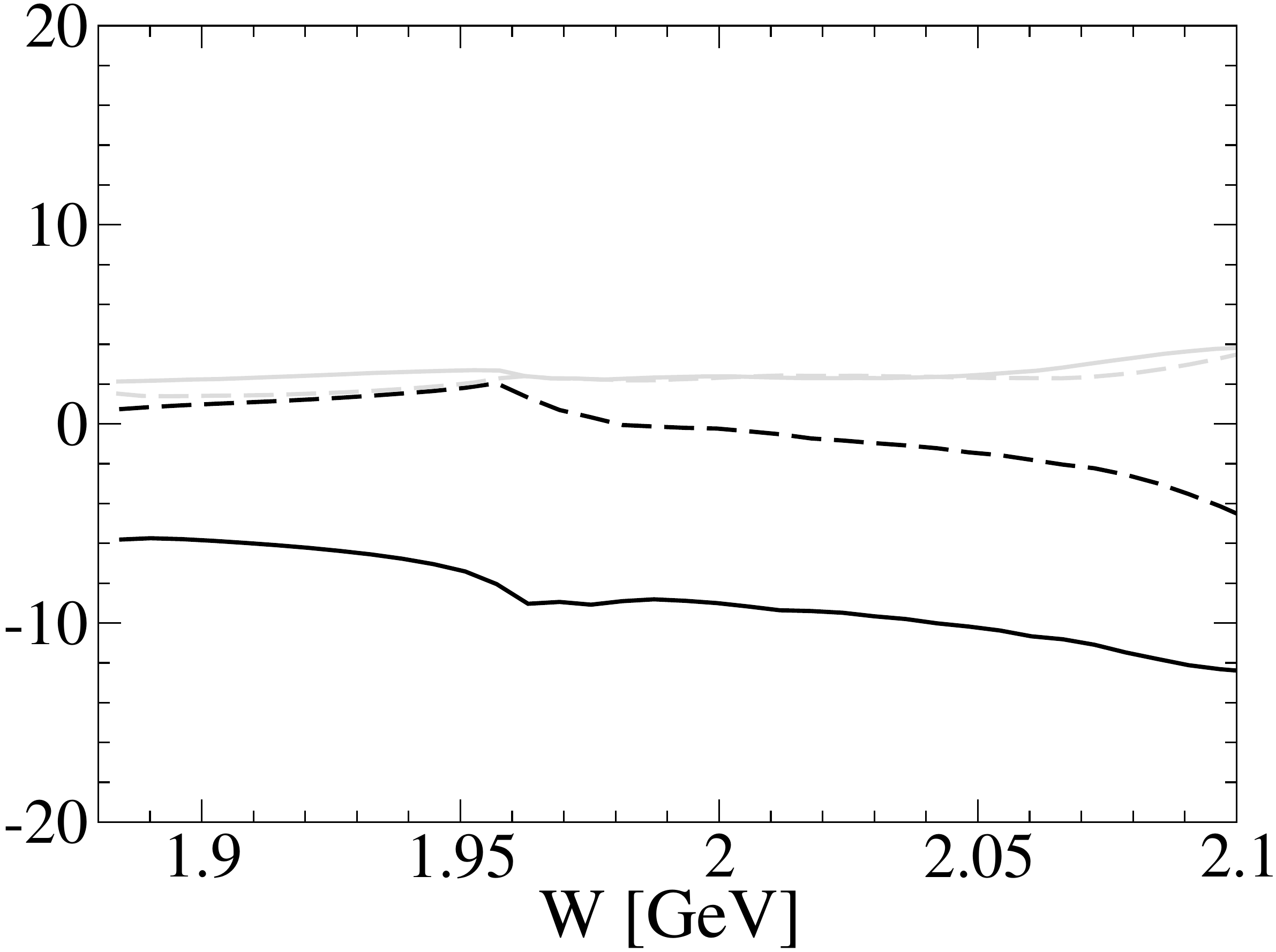}
\includegraphics[width=5.05cm]{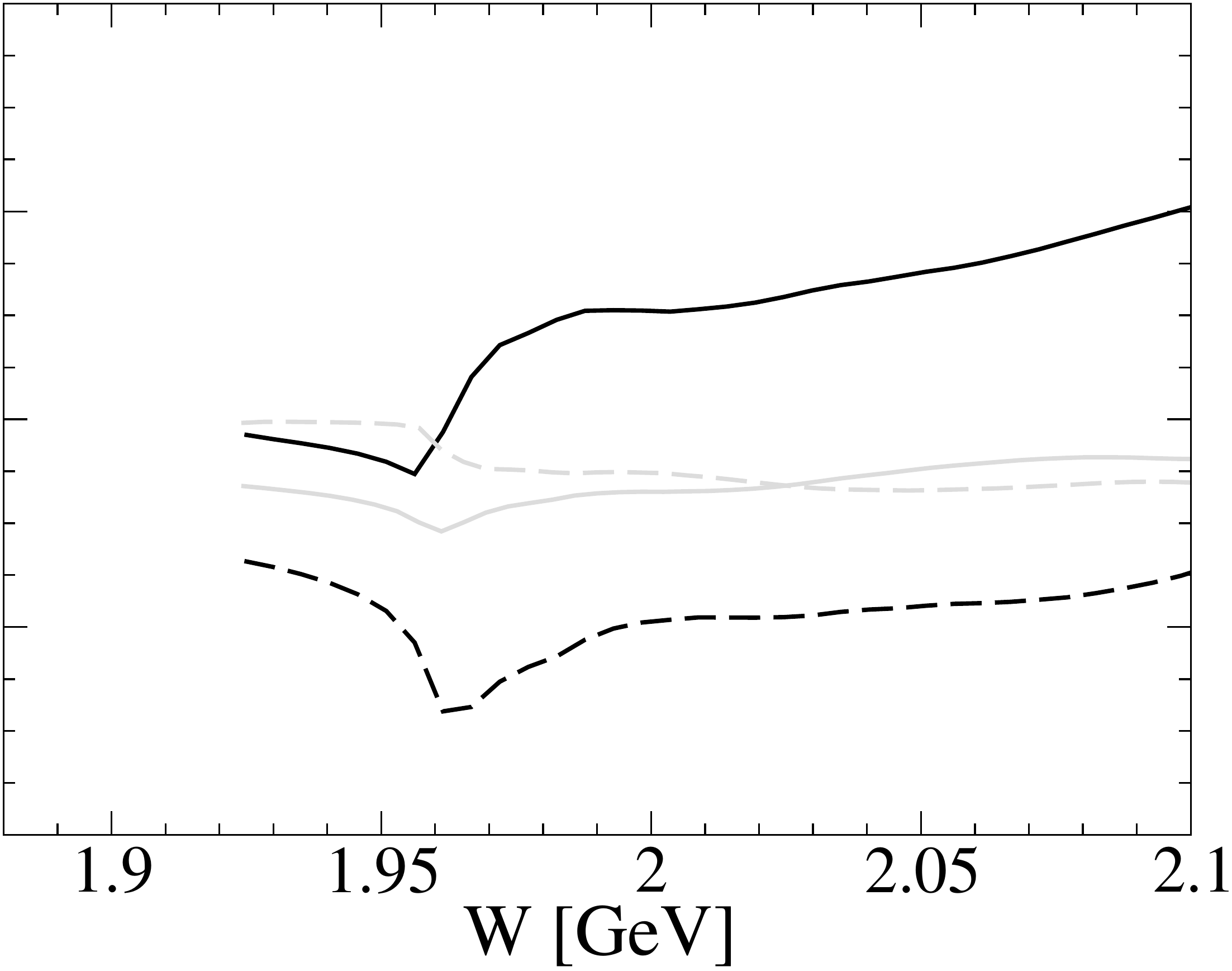}
\includegraphics[width=5.3cm]{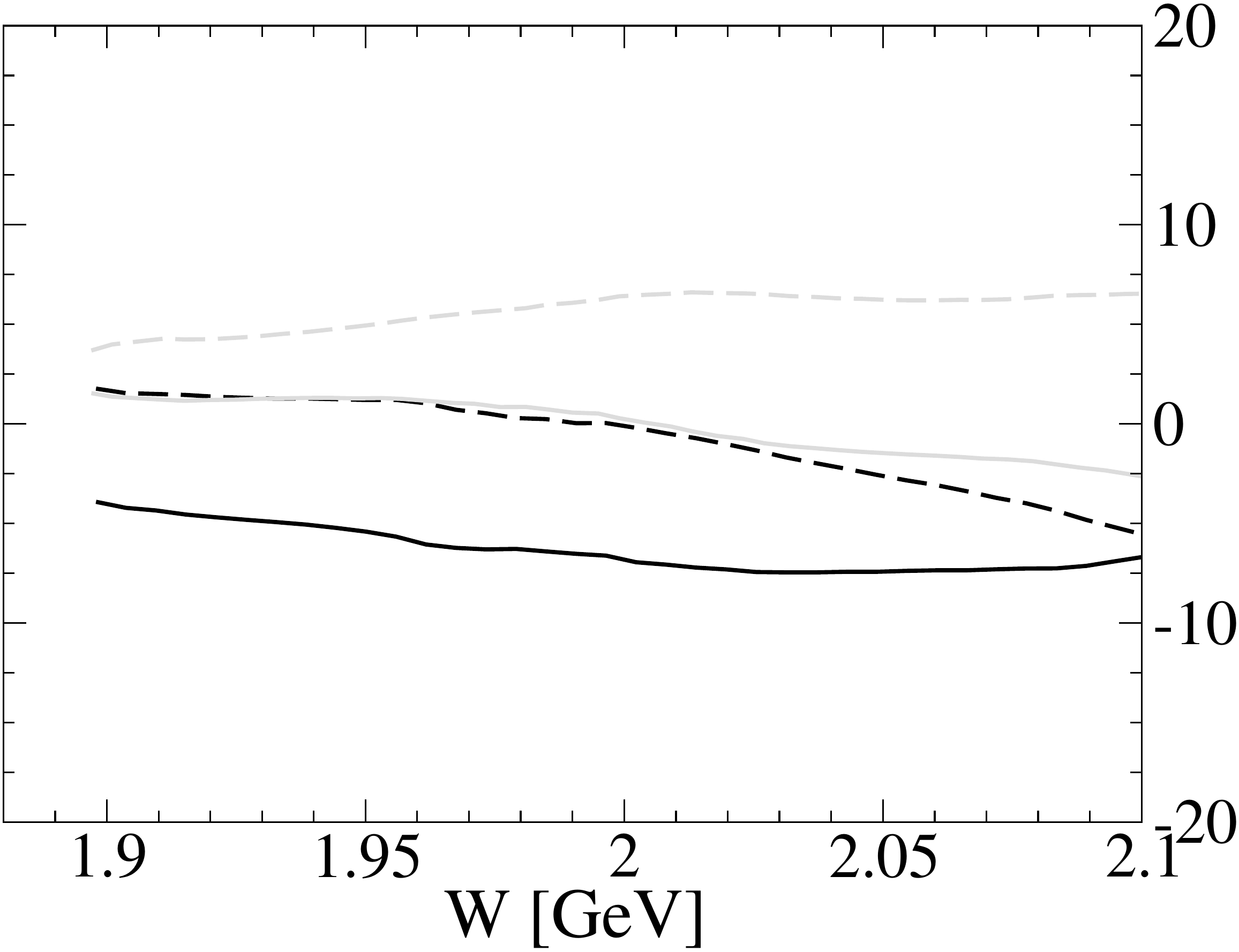}
\caption{Real (solid) and imaginary (dashed) parts of the coupling constants for
(a) $N_{1,2}^{*+} \to K^+\Lambda_1$, (b) $N_{1,2}^{*+} \to K^+\Lambda_2$, and
(c) $N_{1,2}^{*+} \to K^+ \Sigma(1400)$.
The values corresponding to $N_1^*(1801)$ and $N_2^*(1912)$ are shown in dark and
light curves, respectively.}
\label{fig2}
\end{figure}

Alternatively to using a constant value for the $g_{KY^*N^*}$ coupling  in Eq.~(\ref{eq:NsSt}), we can extract their energy dependent values using the $N^*(1895)\to K Y^*$ amplitudes provided in Ref.~\cite{Khemchandani:2020exc}. In this former work, keeping in mind that $N^*(1895)$ has a large width, the amplitude for its decay to hyperon resonances $Y^*$ were determined as a function of  the variable mass of $N^*(1895)$. The mass values range  within the integration limits in Eq.~(\ref{conv}). We determine the energy dependent values of $g_{KY^*N^*}$ using the Lagrangian $\mathcal L_{KY^*N^*}^{1/2^-}$ of Eq.~(\ref{eq:NsSt}), so as to reproduce the $N^*(1895)\to K Y^*$ amplitudes of Ref.~\cite{Khemchandani:2020exc}, and depict them in Fig.~\ref{fig2}.

\subsection{Radiative decays of $N^*(1895)$, $\Lambda(1405)$,
 and $\Sigma(1400)$}
We now need to discuss the EM vertex which can be described in terms of the
effective Lagrangian
\begin{align}
\mathcal L_{\gamma N N^*}^{1/2^-} = \frac{e\mu_{N^*N}}{2M_N} \bar N \gamma_5
\sigma_{\mu\nu} \partial^\nu A^\mu N^* + \mathrm{H.c.},
\label{eq:NsEM}
\end{align}
and $\mathcal L_{\gamma Y Y^*}$ in Eq.~(\ref{eq:Lag:EM}), which are required to calculate the $s$- and $u$-channel diagrams, respectively, in Fig.~\ref{fig1}.
\begin{figure}[h!]
\includegraphics[width=0.3\textwidth]{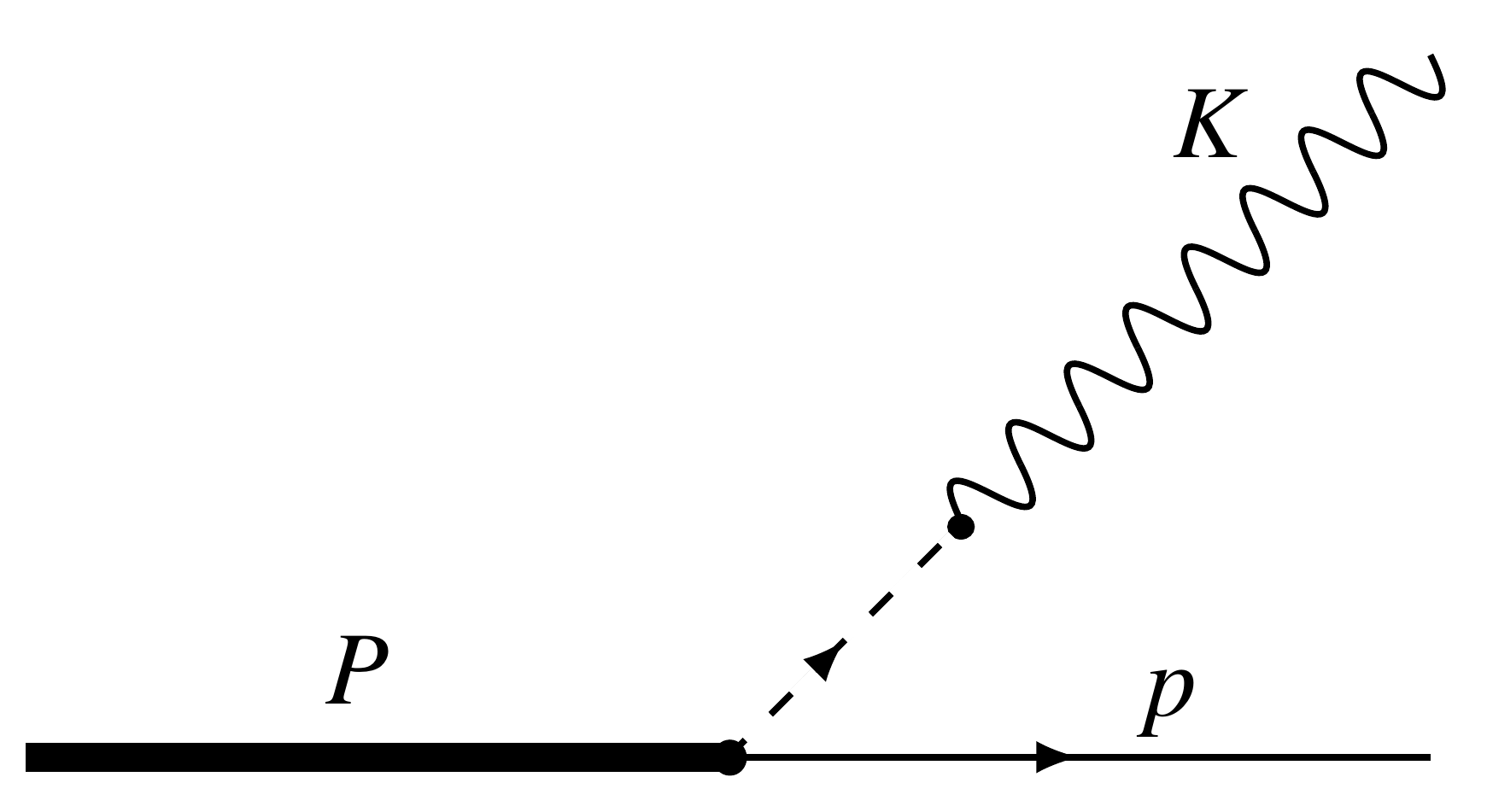}
\caption{Radiative decay of a baryon resonance through VMD. The labels $P$, $K$ and $p$ indicate the four-momenta of the corresponding particles.}\label{fig3}
\end{figure}
To estimate the transition magnetic moments of $N^*(1895)$ as well as of $\Lambda(1405)$ and $\Sigma(1400)$, we employ the vector-meson dominance (VMD) model as shown in Fig.~\ref{fig3}, where the dashed line corresponds to $\rho^0$, $\omega$ or $\phi$.
By using the vector meson-photon vertex Lagrangian
\begin{align}
\mathcal{L}_{V\gamma} =-\frac{e F_V}{2} \lambda_{V\gamma} V_{\mu \nu} A^{\mu \nu},
\label{vgamma}
\end{align}
we can evaluate the diagram of Fig.~\ref{fig3} since the couplings of different vector meson-baryon channels to  $N^*(1895)$, $\Lambda(1405)$, and $\Sigma(1400)$ have already been determined in Refs.~\cite{Khemchandani:2013nma,Khemchandani:2018amu}.
Let us see in more detail the evaluation of the amplitude for the diagram shown in Fig.~\ref{fig3}.
In Eq.~(\ref{vgamma}), $F_V$ is the decay constant for vector mesons, which we take as 154 MeV~\cite{Roca:2003uk}, $A^{\mu \nu}=\partial^\mu A^\nu-\partial^\nu A^\mu$ and $V_{\mu \nu}$ is a tensor field related to $\rho^0$, $\omega$, $\phi$, with $\lambda_ {V\gamma} = 1,~\frac{1}{3},~-\frac{\sqrt{2}}{3}$, respectively. As in Ref.~\cite{Roca:2003uk}, the tensor field $V_{\mu \nu}$ is normalized as
\begin{align}
V^{\mu \nu}= \frac{1}{M_V}\left(\partial^\mu V^\nu-\partial^\nu V^\mu\right),
\end{align}
and the corresponding propagator is written by
\begin{align}
i G_{\alpha\beta\mu\nu}\left(K\right) =\frac{i}{M_V^2\left(M_V^2-K^2\right)}\left(g_{\alpha\mu}g_{\beta\nu}(M_V^2-K^2)+g_{\alpha\mu}K_\beta K_\nu-g_{\alpha\nu}K_\beta K_\mu - \alpha \leftrightarrow \beta\right).
\end{align}
For the VBB$^*$ vertex, we write the effective Lagrangian as
\begin{align}
\mathcal{L}_{VBB^*}=-\tilde g_{VBB^*} \bar B \gamma_5 \sigma_{\mu\nu} B^* V^{\mu\nu\dagger} + \mathrm{H.c}.\label{VBBstr}
\end{align}
This Lagrangian leads to the following amplitude for the VB$\to$ B$^*\to$VB process, in the nonrelativistic limit, 
\begin{align}
t_{V_1B\to V_2B}&=\frac{\tilde g_{V_1BB^*} \tilde g_{V_2BB^*}}{\sqrt{s}-M_{B^*}+i\Gamma/2}~ \frac{4K_1^0 K_2^0}{M_{V1}M_{V2}}~\vec \sigma\cdot\vec\epsilon_2 ~\vec\sigma\cdot\vec \epsilon_1,\label{tbw}
\end{align}
where $M_{V1}$ ($K_1^0$) and $M_{V2}$ ($K_2^0$) denote the mass (energy) of the vector meson in the initial and final state, respectively.
Comparing Eq.~(\ref{tbw}) with the Breit-Wigner parameterization of the amplitudes projected on spin 1/2 in Refs.~\cite{Khemchandani:2018amu,Khemchandani:2013nma}, 
\begin{align}
t_{V_1B\to V_2B}&=\frac{g_{V_1BB^*}g_{V_2BB^*}}{\sqrt{s}-M_{B^*}+i\Gamma_{B^*}/2},
\end{align}
we find the relation between $\tilde g_{V_iBB^*}$ in Eq.~(\ref{VBBstr}) and $g_{V_iBB^*}$ obtained in Refs.~\cite{Khemchandani:2018amu,Khemchandani:2013nma} as
\begin{align}
\tilde g_{V_iBB^*}= \frac{g_{V_iBB^*}}{2\sqrt{3}}\frac{M_{Vi}}{K_i^0},\label{gnorm}
\end{align}
where  $K_i^0$ is the energy of the meson, at the VBB$^*$ vertex, at the resonance mass
\begin{align}
K_i^0 = \frac{M_{B^*}^2+M_{Vi}^2-M_B^2}{2 M_{B^*}}.
\end{align}

Finally, we obtain the amplitude for the $B^* \to B \gamma$ process, using Eqs.~(\ref{vgamma}),~(\ref{VBBstr}), and~ (\ref{gnorm}) as
\begin{align}
t_{B^*\to B \gamma} = \frac{2 e F_V \tilde g_{VBB^*} \lambda_{V\gamma}}{ M_V^2} \bar B \gamma_5 \slashed \epsilon \slashed K B^*,\label{radamp}
\end{align}
with $\epsilon$ denoting the polarization vector for the photon. Notice that the above amplitude is manifestly gauge invariant. 

The radiative decay width can be obtained through
\begin{align}
\Gamma_{B^* \to B \gamma} = \frac{1}{8 \pi^2}\frac{|\,\vec K\,|   M_{B}}{M_{B^*}}\frac{1}{2 S_{B^*} +1}\int d\Omega \sum\limits_{m_{B^*}, m_{B}, m_\gamma} |\mathcal{M}_{B^* \to B \gamma}|^2,\label{eq:width2}
\end{align}
and, using the amplitude in Eq.~(\ref{radamp}), we get
\begin{align}
\Gamma_{B^* \to B \gamma} =\frac{4 e^2 F_V^2  |\vec K|^3}{\pi M_V^4}\Big|\tilde g_{\rho^0BB^*}+\frac{\tilde g_{\omega BB^*}}{3}-\frac{\sqrt{2}\tilde g_{\phi BB^*}}{3}\Big|^2\left(\frac{E_B+M_B}{2M_B}\right),
\end{align}
with $M_V = 770$ MeV, and $\tilde g_{\rho^0 B B^*}$ is related to the $\rho BB^*$ coupling $\tilde g_{\rho B B^*}$ given  in the isospin base by $\tilde g_{\rho^0BB^*}=-\tilde g_{\rho BB^*}/\sqrt{3}$. The  decay widths  determined in this way for $\Lambda(1405)$, $\Sigma(1400)$ and $N^*(1895)$ are given in Table~\ref{TAB3}. It should be noted that these results have been obtained by convoluting the partial widths over the finite widths of $B^*$ as in Eq.~(\ref{conv}).
\begin{table}[h]
\begin{tabular}{cc||cc}
\hline\hline
Decay process& Partial width (KeV)&Decay process& Partial width (KeV) \\ 
\hline
$\Lambda_1(1405) \to \Lambda \gamma$&9.47 $\pm$ 2.17&$N_1^*(1895) \to p \gamma$&729.17 $\pm$ 78.20\\
$\Lambda_2(1405) \to \Lambda \gamma$&11.91 $\pm$ 3.39&$N_2^*(1895) \to p \gamma$&129.59 $\pm$ 13.89\\
\underline {$\Lambda(1405) \to \Lambda \gamma$}&26.19 $\pm$ 6.93&\underline {$N^*(1895) \to p \gamma$}&650.70 $\pm$ 65.10\\
$\Lambda_1(1405) \to \Sigma \gamma$&5.17 $\pm$ 1.75&$\Sigma(1400) \to \Lambda \gamma$&49.97 $\pm$ 8.57\\
$\Lambda_2(1405) \to \Sigma \gamma$&2.08 $\pm$ 1.72&$\Sigma(1400) \to \Sigma \gamma$&94.51$ \pm$ 9.33\\
$\Lambda(1405) \to \Sigma \gamma$&2.50 $\pm$ 1.37&&\\
\hline\hline
\end{tabular}
\caption{Radiative decay widths for $\Lambda(1405)$, $\Sigma(1400)$ and $N^*(1895)$. The underlined process means that an interference between the two poles related to the decaying hadron has been considered to obtain the decay width.}
\label{TAB3}
\end{table}

An important point to consider is that  $\Lambda(1405)$ as well as $N^*(1895)$ are related to two poles in the complex plane. We provide the decay widths for each pole of $\Lambda(1405)$ and $N^*(1895)$ separately. However, since in each case, the poles have overlapping widths and experimentally they  may be observed as one state, we find it  useful to obtain the decay width of such a state. To do this we recall that the effect of two close lying poles in the complex plane, on the real axis, can be produced by an interference between meson-baryon amplitudes related to the two poles written  within the Breit-Wigner description. In the same spirit, we calculate the decay width by allowing an interference between the couplings of the two poles and by using a mass and width which is an average value obtained from the pole positions. Such decay widths are underlined in Table.~\ref{TAB3}.

 We can now compare our results with the information available from the experiments. The radiative decay width of $\Lambda(1405)\to\Lambda \gamma$ determined from the experimental data is known to be  $27\pm8$~KeV~\cite{Zyla:2020zbs}. Our result obtained by considering the superposition of the two poles is in remarkable agreement with the experimental data. For $\Lambda(1405)\to\Sigma \gamma$, PDG~\cite{Zyla:2020zbs} provides two possible values: $10\pm4$~KeV or $23\pm7$~KeV. Our results are closer to the former value.

In case of $N^*(1895)$, the branching ratio of the radiative decay is known to be 0.01-0.06$\%$~\cite{Zyla:2020zbs}. In our case, the branching ratio for $N_1^*$ is 0.34$-$0.42$~\%$, while for $N^*_2$ is 0.11$-$0.13$~\%$. If we consider the superposition of the two poles of $N^*(1895)$, which produces a peak on the real axis with an average width of about 120 MeV, we obtain a branching ratio $\sim$0.49$-$0.60$~\%$. Our results for the second pole seem to be closer  to the upper limit of the value listed in Ref.~\cite{Zyla:2020zbs}. Actually the real and imaginary part  of this second pole are closer to the values associated with $N^*(1895)$ in Ref.~\cite{Zyla:2020zbs}.  Further, it should be mentioned that making a comparison is difficult in this case, since all structures above 1800 MeV appearing in the $S_{11}$ wave are listed under the label of $N^*(1895)$ in Ref.~\cite{Zyla:2020zbs}. And due to this former fact, information from different states might be associated with $N^*(1895)$. In fact, the branching ratios are estimated in Ref.~\cite{Zyla:2020zbs} by using helicity amplitudes from Refs.~\cite{Hunt:2018wqz,Sokhoyan:2015fra}, where the former work associates a pole of $1956-i449/2$ MeV with  $N^*(1895)$ while the latter one finds $\left(1907 \pm 10\right) -i\left(100^{+40}_{-15}\right)/2$ MeV.

Having the decay widths in Table~\ref{TAB3}, we obtain the transition magnetic moments related to each decay using the relation
\begin{align}
\Gamma_{B^* \to B \gamma}= \frac{\left(e\mu_{B^* B}\right)^2 |\vec K|^3}{4 \pi M_N^2},
\end{align}
and their values are summarized in Table~\ref{TAB4}.
\begin{table}[h]
\begin{tabular}{cc||cc}
\hline\hline
Decay process& Magnetic moment&Decay process& Magnetic moment \\ 
\hline
$\Lambda_1(1405) \to \Lambda \gamma$&0.28 $\pm$ 0.02&$N_1^*(1895) \to p \gamma$&0.56 $\pm$ 0.02\\
$\Lambda_2(1405) \to \Lambda \gamma$&0.26 $\pm$ 0.02&$N_2^*(1895) \to p \gamma$&0.20 $\pm$ 0.01\\
\underline {$\Lambda(1405) \to \Lambda \gamma$}&0.42 $\pm$ 0.03&\underline {$N^*(1895) \to p \gamma$}&0.45 $\pm$ 0.02\\
$\Lambda_1(1405) \to \Sigma \gamma$&0.33 $\pm$ 0.03&$\Sigma(1400) \to \Lambda \gamma$&0.60 $\pm$ 0.03\\
$\Lambda_2(1405) \to \Sigma \gamma$&0.15 $\pm$ 0.04&$\Sigma(1400) \to \Sigma \gamma$&1.28 $\pm$ 0.04\\
$\Lambda(1405) \to \Sigma \gamma$&0.20 $\pm$ 0.03&&\\
\hline\hline
\end{tabular}
\caption{Transition magentic moments related to decays of $\Lambda(1405)$, $\Sigma(1400)$ and $N^*(1895)$. The underlined process means that a superposition of the two poles associated with the decaying hadron has been considered to obtain the decay width.}
\label{TAB4}
\end{table}
The transition magnetic moments $\mu_{\Lambda^*\Lambda}$ determined in our work for the individual poles of $\Lambda(1405)$, as well as for their superposition, are compatible with those obtained within the chiral unitary model of Ref.~\cite{Jido:2002yz}.

\section{Numerical results and Discussions}
\label{SecIII}
Let us show and discuss our numerical results.
We first reproduce the $\gamma p \to K^+ \Lambda^*$ reaction and use the same
model parameters for predicting the  observables for the $\gamma p \to K^+
\Sigma^*$ process except for some coupling constants which are determined in
Sec.~\ref{SecII}.
The cutoff masses in Eq.~(\ref{eq:FF}) and~(\ref{eq:GauFF}) are determined to
be $\Lambda_{N,\Lambda,\Sigma} =0.9$~GeV and $\Lambda_{N^*} =
0.83$ GeV, respectively.
\begin{figure}[htp]
\centering
\includegraphics[width=7.0cm]{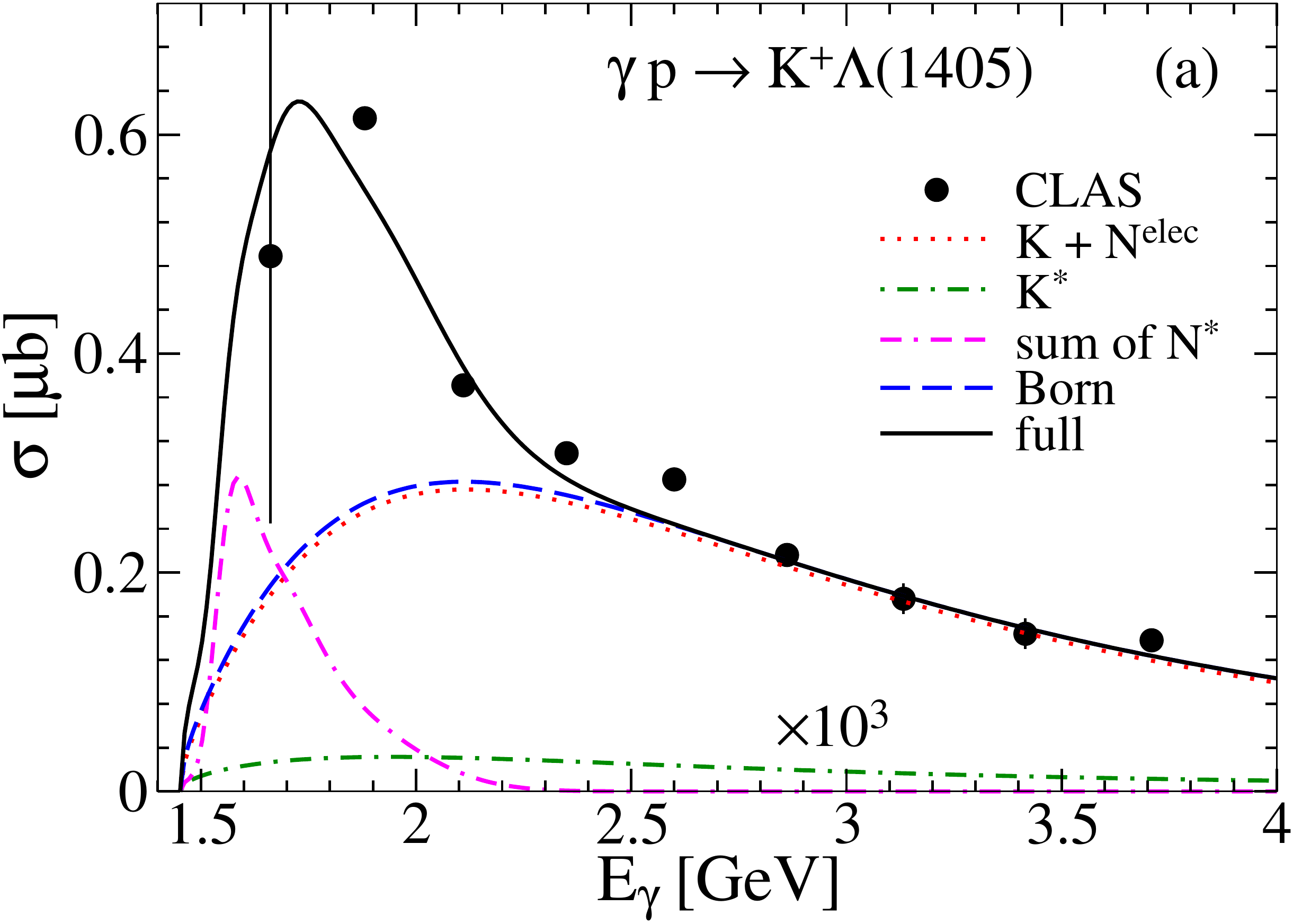} \hspace{0.3em}
\includegraphics[width=7.0cm]{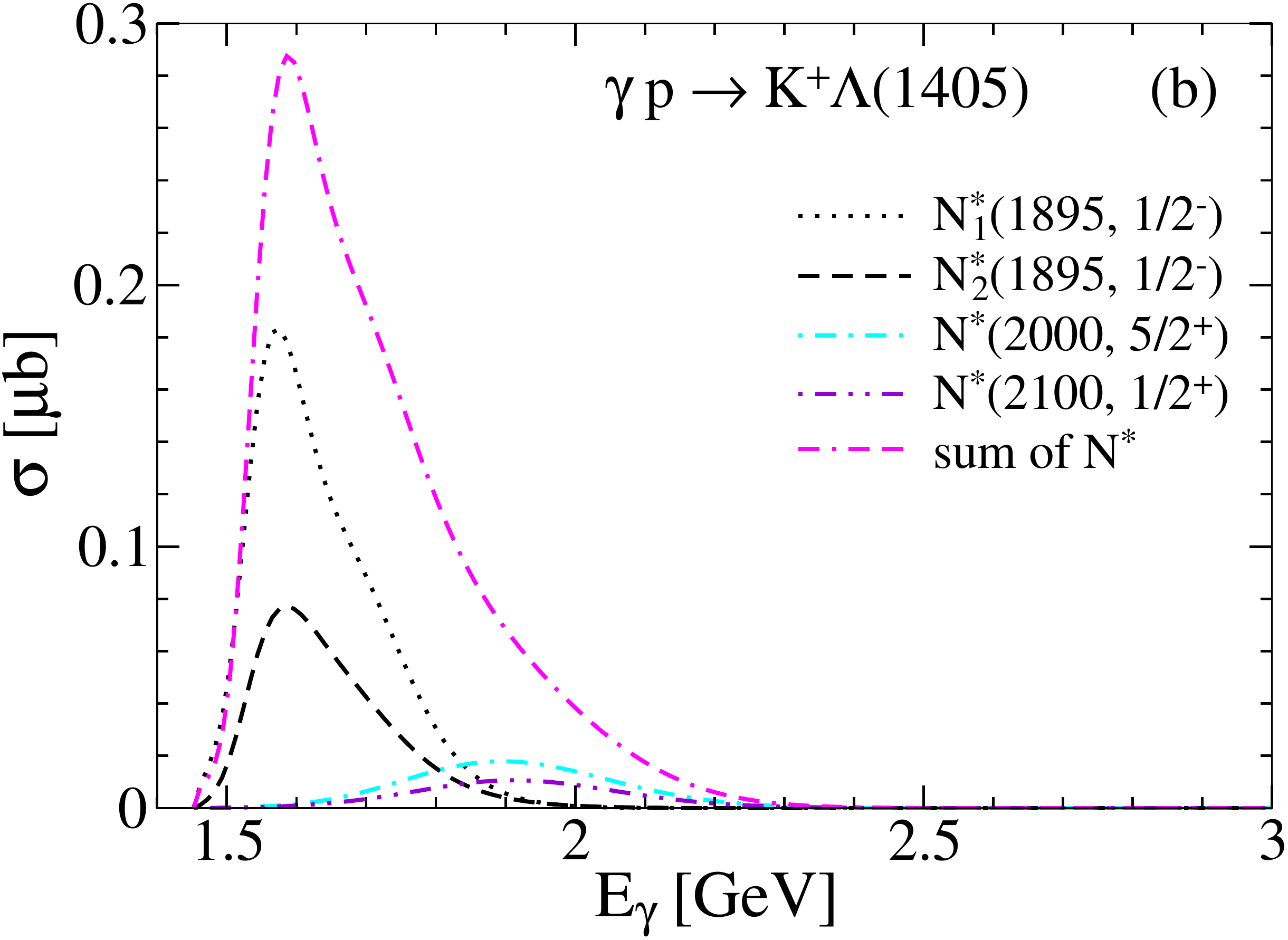}
\caption{(a) Total cross section for $\gamma p \to K^+ \Lambda(1405)$ is plotted
as a function of the lab energy $E_\gamma$.
The red dotted curve represents the sum of the $K$-Reggeon and electric part of
$N$ contributions.
The green dot-dashed and the magenta dot-dashed-dashed curves denote the
$K^*$-Reggeon and the $N^*$ contributions, respectively.
The blue dashed and the black solid curves stand for the Born-term and the
full contributions, respectively.
(b) Each of the $N^*$ contributions is plotted.
(a) The data are taken from the CLAS Collaboration~\cite{Moriya:2013hwg}.
The $K^*$-Reggeon contribution is multiplied by the factor of $10^3$ for easy
comparison.}
\label{fig4}
\end{figure}
\begin{figure}[htp]
\centering
\includegraphics[width=7.0cm]{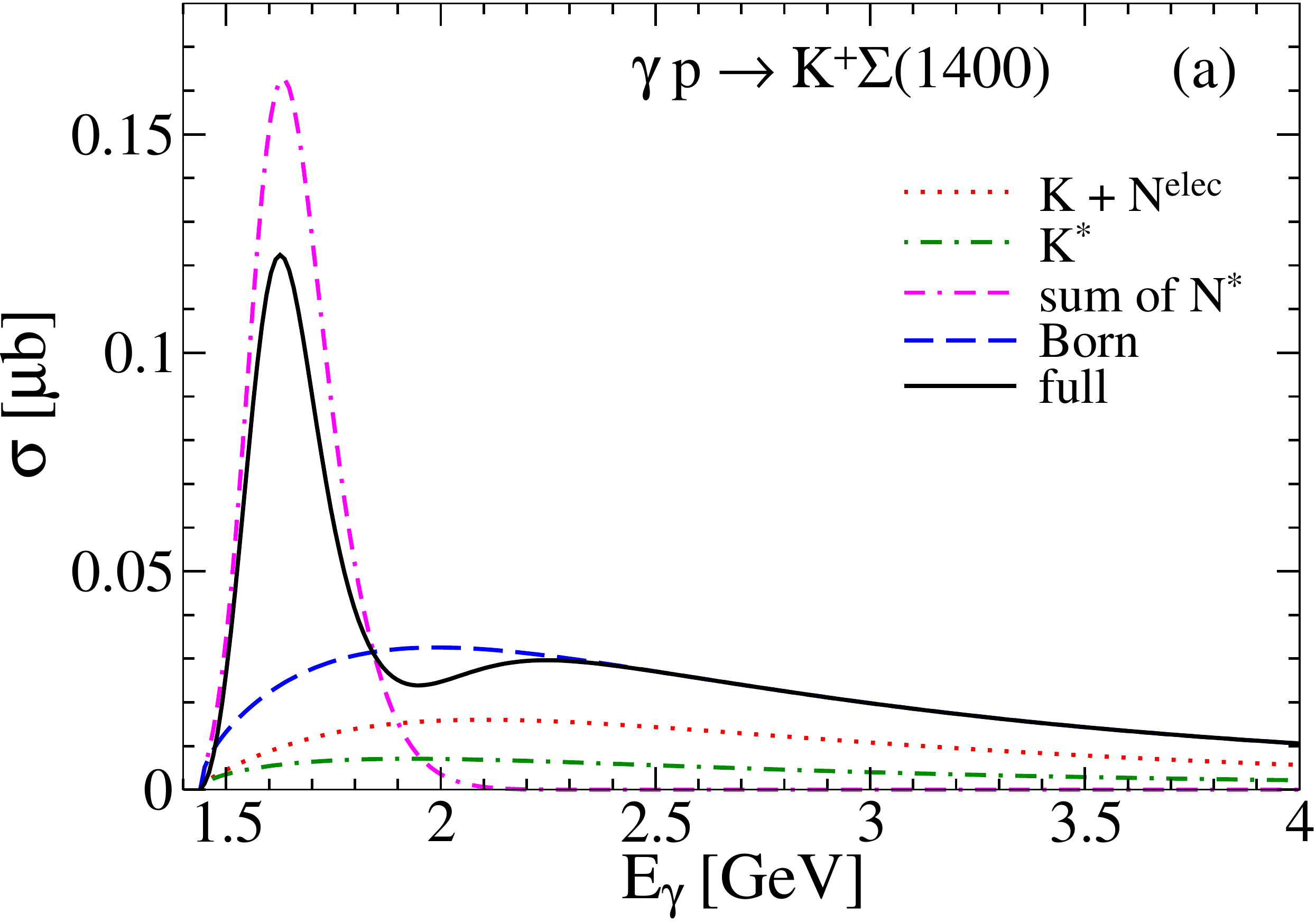} \hspace{0.3em}
\includegraphics[width=7.0cm]{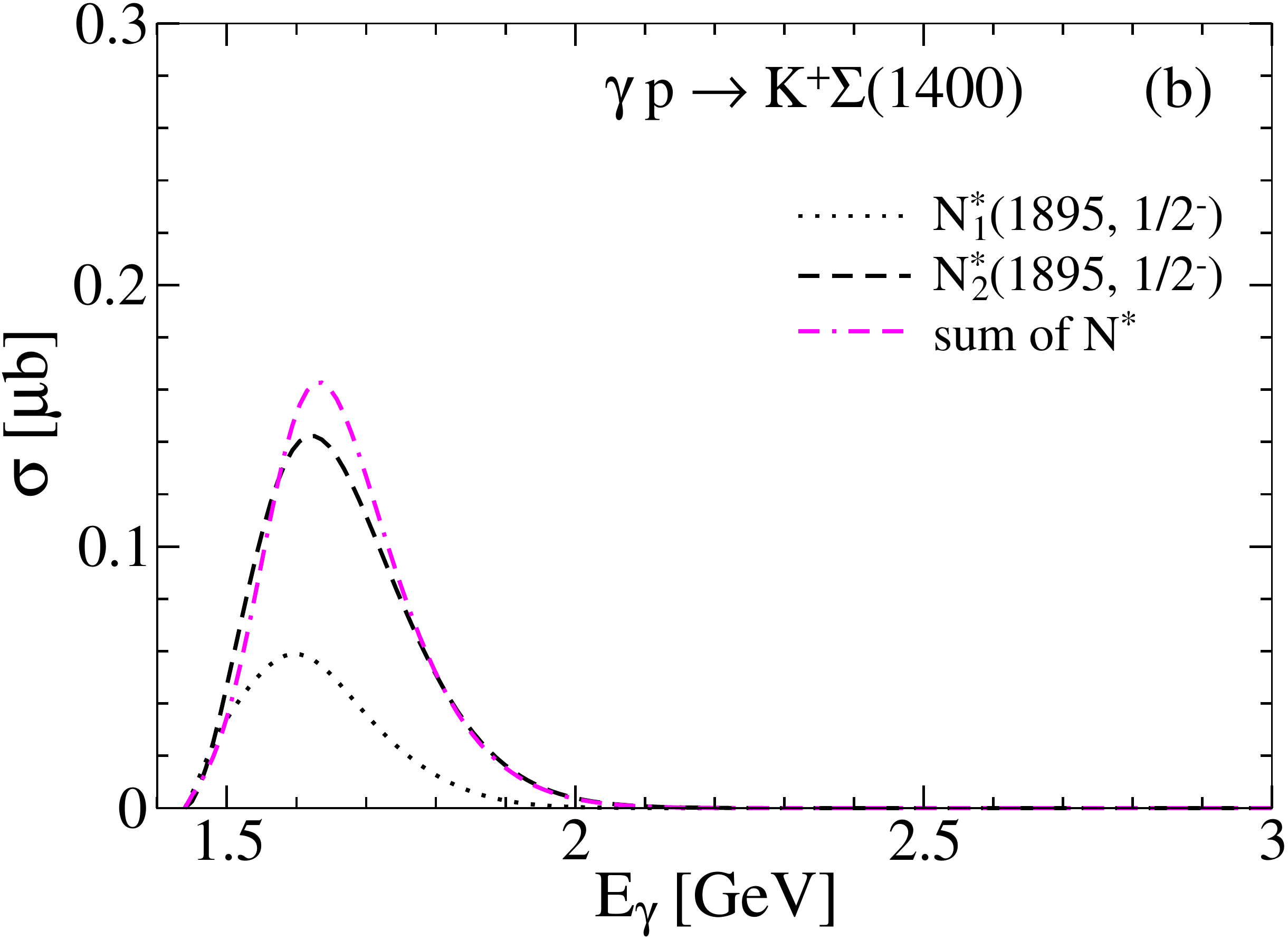}
\caption{(a) Total cross section for $\gamma p \to K^+ \Sigma(1400)$ is plotted
as a function of the lab energy $E_\gamma$ .
The curve notations are the same as Fig.~\ref{fig4}(a).
(b) Each of the $N^*$ contributions is plotted.}
\label{fig5}
\end{figure}
The total cross section for $\gamma p \to K^+ \Lambda^*$ is displayed as a
function of the photon laboratory (lab) energy $E_\gamma$ in Fig.~\ref{fig4}(a).
It turns out that the constant $K$ Regge phase in Eq.~(\ref{eq:ReggeProp})
produces good results regardless of the $K^*$ phase.
The CLAS data~\cite{Moriya:2013hwg} at lab energies above 2.5 GeV are reproduced
quite well by the Born-term contribution mostly due to the $K$-Reggeon exchange.
The contribution of the $K^*$-Reggeon exchange is highly suppressed because of
the small value of $g_{K^*N\Lambda^*}$ relative to $g_{KN\Lambda^*}$.
The low-energy region ($E_\gamma \leqslant 2.5$ GeV) is matched after we
additionally include the $N^*$ contributions, each of them is depicted in
Fig.~\ref{fig4}(b).
The previous study~\cite{Kim:2017nxg} included two PDG resonances,
$N^*(2000,\,5/2^+)$ and $N^*(2100,\,1/2^+)$, and three missing resonances,
$N^*(2030,1/2^-)$, $N^*(2055,\,3/2^-)$, and $N^*(2095,\,3/2^-)$.
It attributed a major role to the two PDG resonances.
In this work, we additionally include the $N^*(1895)$ that has a two pole nature
as discussed in Sec.~\ref{SecII}.
The larger discrepancy between the Born-term contribution and the CLAS data at
$E_\gamma \leqslant 2.5$ GeV as compared to the results in
Ref.~\cite{Kim:2017nxg} is due to the small modification in constructing a
gauge-invariant amplitude (See Eq.~(\ref{eq:Amp:SRegge})).
It is interesting that including $N^*(1895)$ provides a satisfactory description
of $\gamma p \to K^+ \Lambda^*$.
The two poles of $N^*(1895)$ interfere constructively and their sum reaches
around 0.3 $\mathrm{\mu b}$ at $E_\gamma = 1.6$ GeV as shown in
Fig.~\ref{fig4}(b).

We present our prediction of the total cross section for $\gamma p \to K^+
\Sigma^*$ in Fig.~\ref{fig5}(a).
The cross section attains a maximum value of about 0.12 $\mathrm{\mu b}$ at
$E_\gamma = 1.6$ GeV.
This magnitude is large enough to be measured in the future experiments.
We find that the contribution from the Born term is nearly an order of magnitude
smaller than that in $\gamma p \to K^+ \Lambda^*$.
The contribution of the $K^*$-Reggeon exchange is comparable to that of the
$K$-Reggeon exchange and even larger than the case of $\gamma p \to K^+
\Lambda^*$.
Here we include the two poles of $N^*(1895)$ as for the $N^*$ contributions
and each of them is contained in Fig.~\ref{fig5}(b).
Although the Born term for $\gamma p \to K^+ \Sigma^*$ is much suppressed
compared to that for $\gamma p \to K^+ \Lambda^*$,
the $N^*$ contribution decreases by  about  50$\%$ only.

\begin{figure}[htp]
\centering
\includegraphics[width=8.6cm]{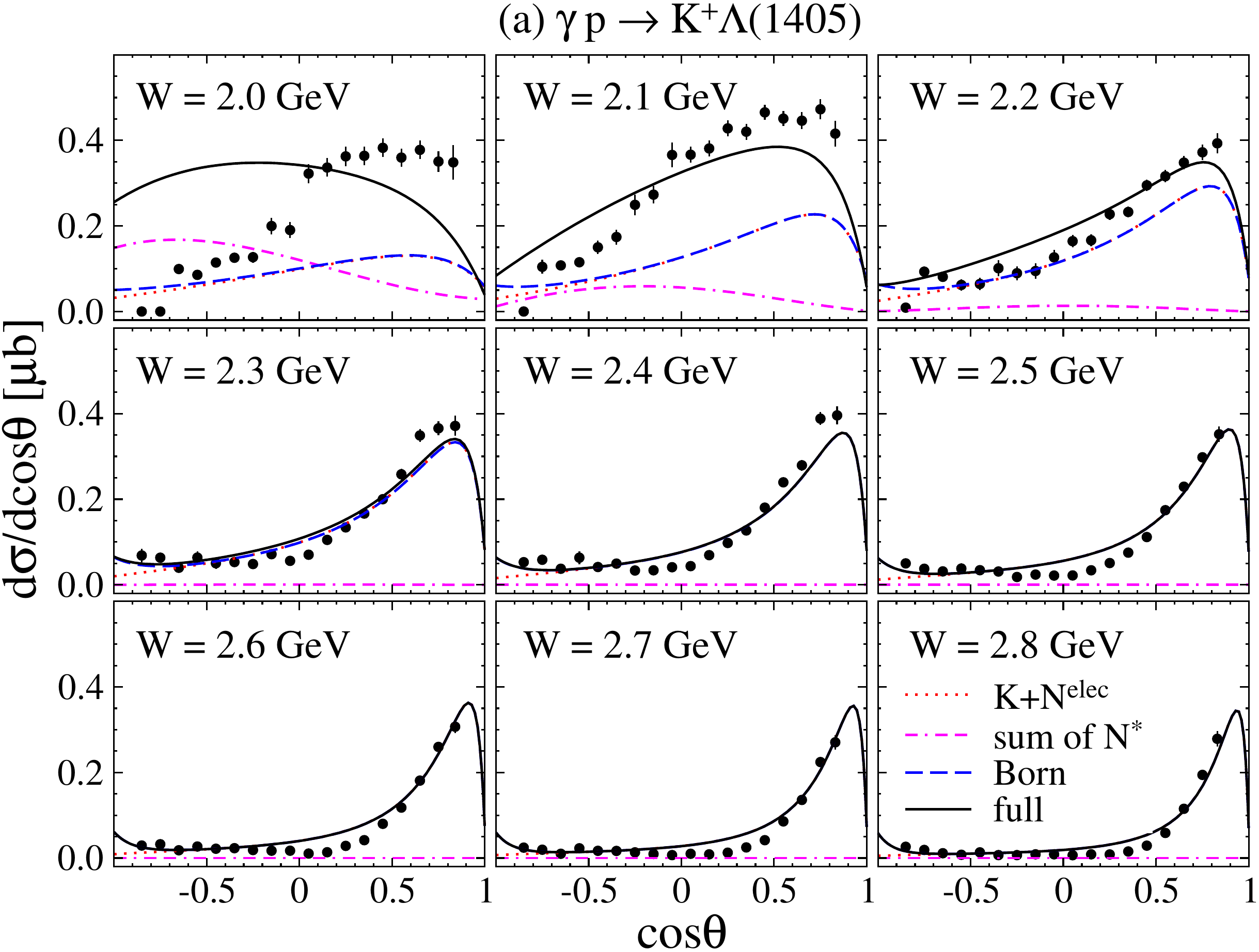} \hspace{0.1em}
\includegraphics[width=8.6cm]{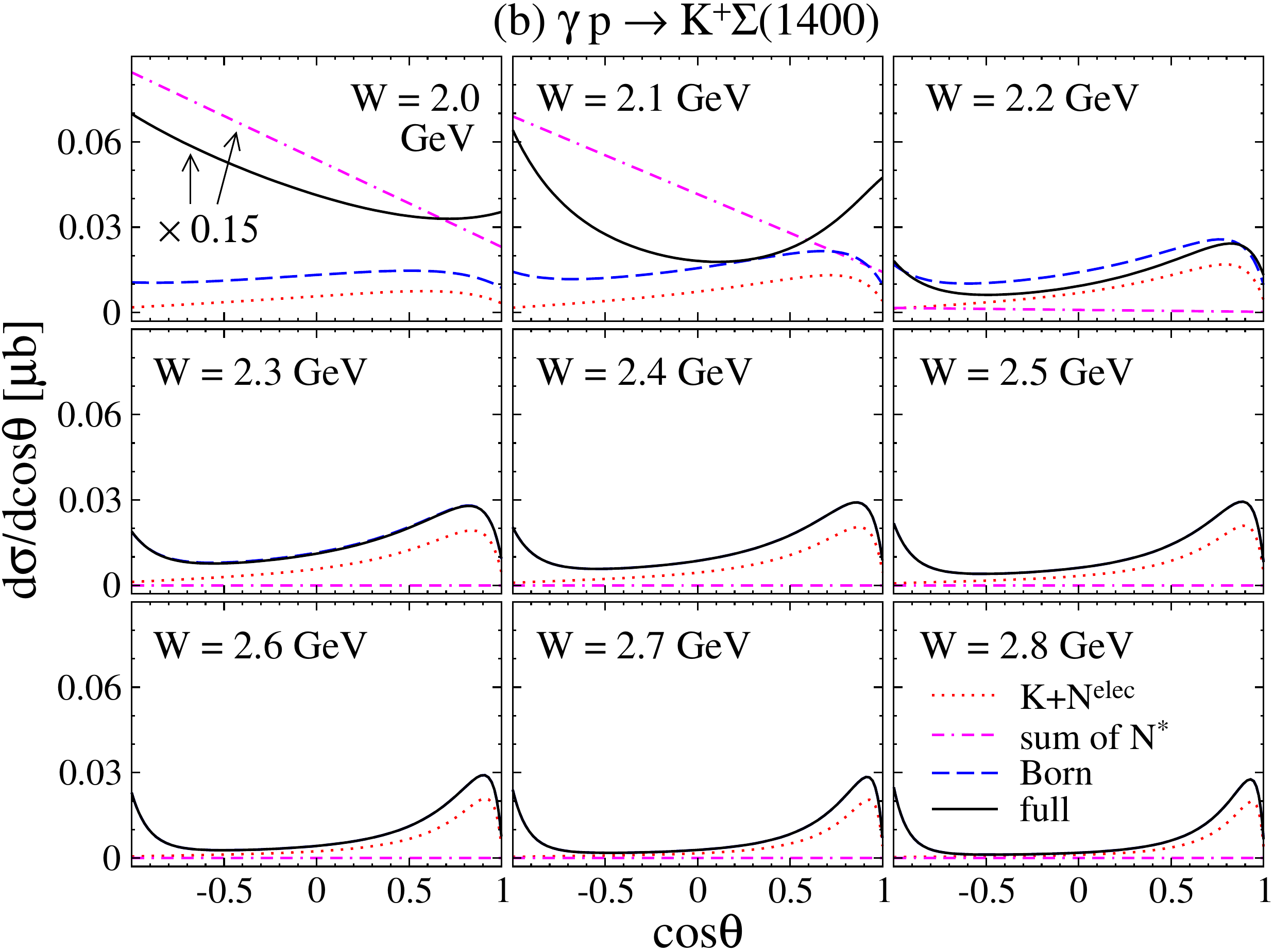}
\caption{Differential cross sections $d\sigma/d\cos\theta$ are plotted as
functions of $\cos\theta$ for (a) $\gamma p \to K^+ \Lambda(1405)$ and (b)
$\gamma p \to K^+ \Sigma(1400)$ for different c.m. energies $W = (2.0 - 2.8)$
GeV.
A constant ($1$) Regge phase for the $K$ and $K^*$ trajectories is used.
The curve notations are the same as Fig.~\ref{fig4}(a).
(a) The data are taken from the CLAS Collaboration~\cite{Moriya:2013hwg}.}
\label{fig6}
\end{figure}
\begin{figure}[htp]
\centering
\includegraphics[width=8.6cm]{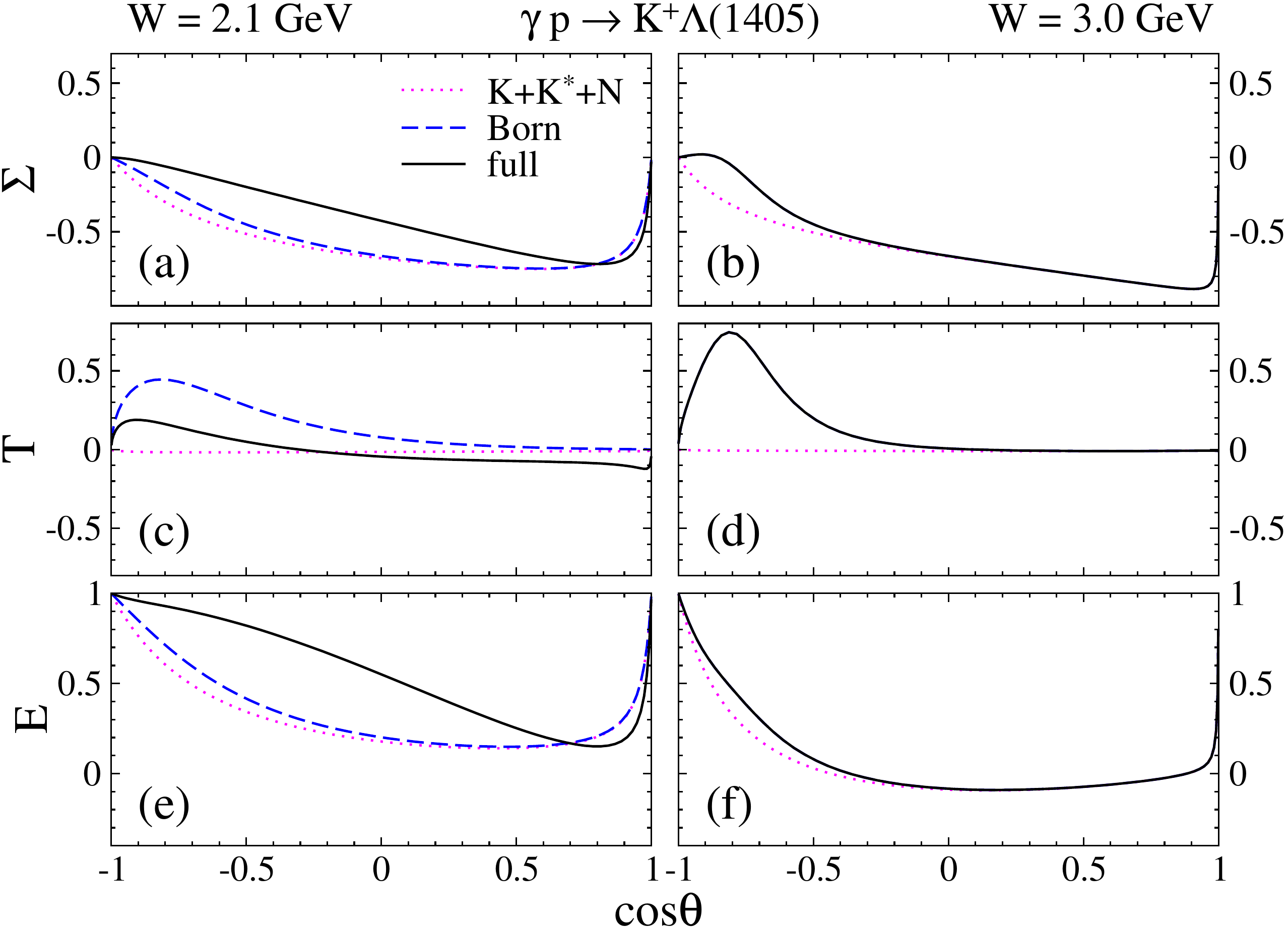} \hspace{0.1em}
\includegraphics[width=8.6cm]{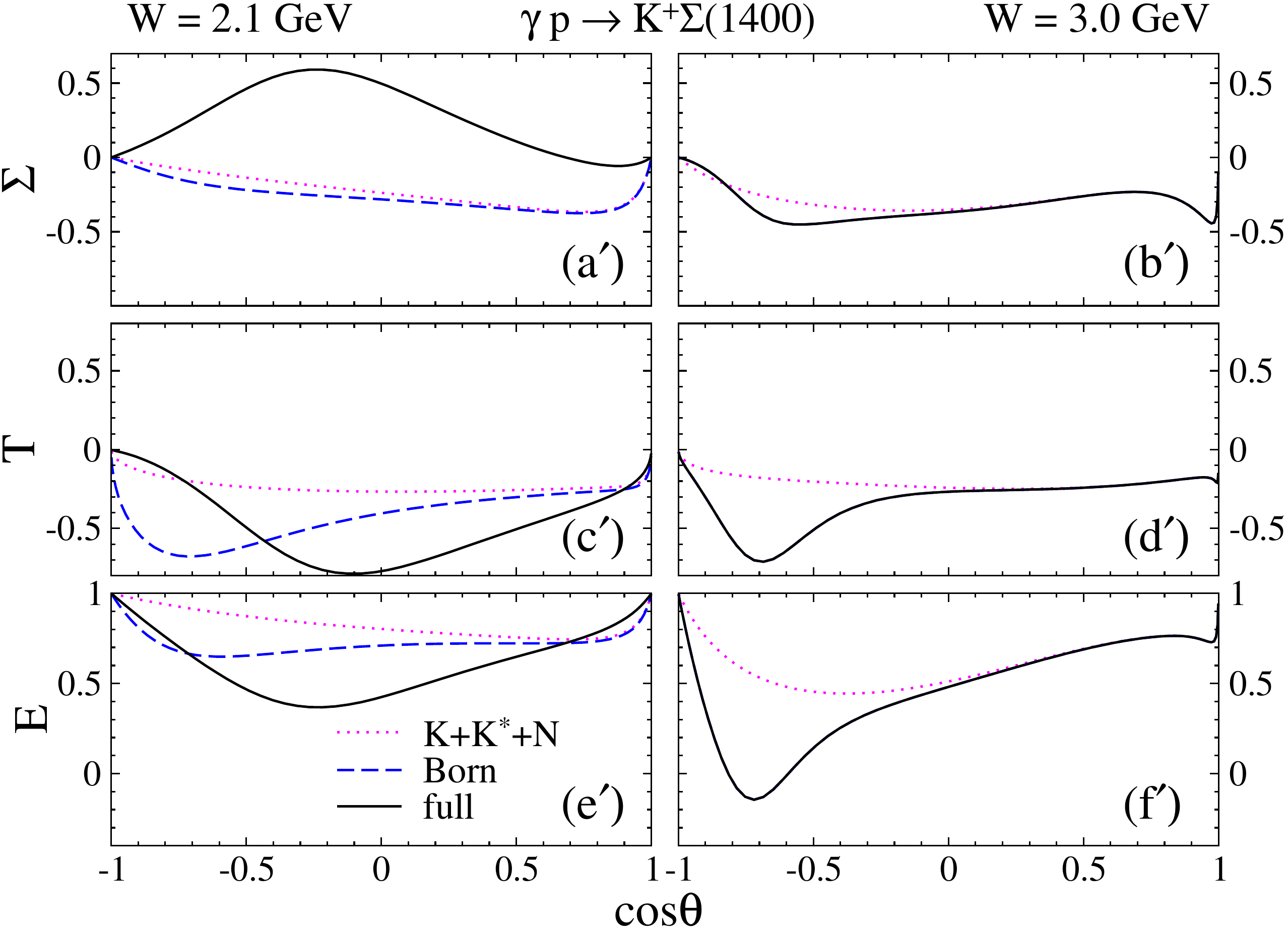}
\caption{Photon-beam asymmetry ($\Sigma$), target asymmetry ($T$), and
beam-target asymmetry ($E$) are plotted as functions of $\cos\theta$ for
$\gamma p \to K^+ \Lambda(1405)$ and $\gamma p \to K^+ \Sigma(1400)$ in the left
and right panels, respectively, at
(a,\,c,\,e,\,a$'$,\,c$'$,\,e$'$) $W$ = 2.1 and (b,\,d,\,f,\,b$'$,\,d$'$,\,f$'$)
3.0 GeV.
The magenta dotted curve denotes the sum of the $K$-Reggeon, $K^*$-Reggon, and
$N$ contributions.
The blue dashed and the black solid curves stand for the Born-term and the
full contributions, respectively.}
\label{fig7}
\end{figure}

Figure~\ref{fig6}(a) depicts the angle dependence of the differential cross
sections for $\gamma p \to K^+ \Lambda^*$.
The $K$-Reggeon exchange shows steadily increasing behavior with $\cos\theta$
and falls off drastically at very forward angles.
The small increase in the backward angle regions at $W \geqslant 2.3$ GeV arises
from the $u$-channel hyperon contributions.
The $N^*$ contributions are mostly due to the two pole structures of
$N^*(1895)$ and become larger as $\cos\theta$ decreases.
The full curves show an excellent agreement with the CLAS data at
$W \geqslant 2.1$ GeV.
Our predictions of the differential cross sections for $\gamma p \to K^+
\Sigma^*$ are displayed in Fig.~\ref{fig6}(b).
The $K$- and $K^*$-Reggon exchanges interfere constructively and mainly consist
of the Born-term contribution.
Both contributions increase gradually as $\cos\theta$, but fall off sharply
at very forward angles because of their spin structures.
The backward peaks are as visible as the forward ones due to the large values of
$\mu_{\Sigma^*\Lambda(\Sigma)}$.

At the level of the unpolarized observables, it is difficult to confirm the
reaction mechanism.
We depict our results of some polarized observables, i.e., the photon-beam
asymmetry ($\Sigma$), the target asymmetry ($T$), and the beam-target asymmetry
($E$) in Fig.~\ref{fig7} as functions of $\cos\theta$ at $W$ = 2.1 and 3.0 GeV.
Their definitions are given by
\begin{align}
\Sigma =& \frac{\sigma^{(\perp,0,0)}-\sigma^{(\parallel,0,0)}}
{\sigma^{(\perp,0,0)}+\sigma^{(\parallel,0,0)}},                                 \cr
T =& \frac{\sigma^{(0,+y,0)}-\sigma^{(0,-y,0)}}
{\sigma^{(0,+y,0)}+\sigma^{(0,-y,0)}},                                         \cr
E =& \frac{\sigma^{(r,+z,0)}-\sigma^{(r,-z,0)}}{\sigma^{(r,+z,0)}+\sigma^{(r,-z,0)}}, \cr
\label{eq:PolObs}
\end{align}
where the superscripts $(B,T,R)$ stand for the polarization states of the photon,
target nucleon, and recoil hyperon, respectively.
We assume that the $\hat z$-axis is the direction of the incident photon beam.
When the photon is polarized in the $\hat x$ and $\hat y$ directions, we use the
labels $\parallel$ (parallel to the reaction plane) and $\perp$ (perpendicular to
the reaction plane), respectively.  In Eq.~(\ref{eq:PolObs}),
$r$  denotes the helicity +1 circular polarization state of the photon.

We first focus on the low-energy region $W = 2.1$ GeV.
As for $\gamma p \to K^+ \Lambda^*$, the negative values of the photon-beam
asymmetry $\Sigma$ for the Born term shown in Fig.~\ref{fig7}(a) are due to the
predominant contribution of the unnatural-parity exchange, i.e., $K$-Reggon
exchange.
Meanwhile, $\Sigma$ for $\gamma p \to K^+ \Sigma^*$ for the Born term is
relatively increased and becomes close to zero at $\cos\theta \leqslant 0$,
as seen in Fig.~\ref{fig7}(a$'$).
This is because the exchange of the natural-parity particle, i.e., the
$K^*$-Reggeon, is as large as that of the $K$-Reggeon and mostly influences 
$\sigma^\perp$.
When the $N^*$  contribution is included to the Born term, $\Sigma$ are increased
for both reactions.

In case of the target asymmetry $T$, we know that the individual particle
exchange is consistent with zero.
As for $\gamma p \to K^+ \Lambda^*$, the Born-term contribution turns out to be
close to zero at $\cos\theta \geqslant 0$, as shown in Fig.~\ref{fig7}(c),
since the single $K$-Reggeon exchange dominates.
The deviation from zero at $\cos\theta \leqslant 0$ arises from the interference
between the $K$-Reggeon and the $u$-channel hyperon exchanges.
Meanwhile, as seen in Fig.~\ref{fig7}(c$'$), $T$ for $\gamma p \to K^+ \Sigma^*$
for the Born term is nonzero and reaches around -0.2 at $\cos\theta \geqslant 0$
because of the interference between the $K$- and $K^*$-Reggon exchanges.
The $u$-channel hyperon contribution is also observed at $\cos\theta \leqslant 0$.
The effect of the $N^*$ exchange is more dramatic in case of
$\gamma p \to K^+ \Sigma^*$.
The results of the beam-target asymmetry $E$ also reveal a distinctive behavior
when we compare both reactions as shown in Figs.~\ref{fig7}(e) and (e$'$).

In a higher energy region, $W$ = 3.0 GeV, the effect of the $N^*$ exchange is
negligible and thus we can easily find how the $u$-channel hyperon exchange comes
into play.
It is interesting that the $u$-channel contribution clearly exhibits different
structures at $\cos\theta \leqslant 0$  for $\Sigma$, $T$,
and $E$ as illustrated in Figs.~\ref{fig7}(b,\,d,\,f,\,b$'$,\,d$'$,\,f$'$).

\begin{figure}[htp]
\centering
\includegraphics[width=7.0cm]{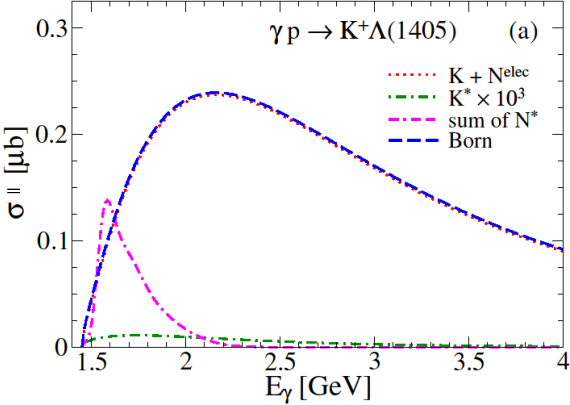} \hspace{0.3em}
\includegraphics[width=7.0cm]{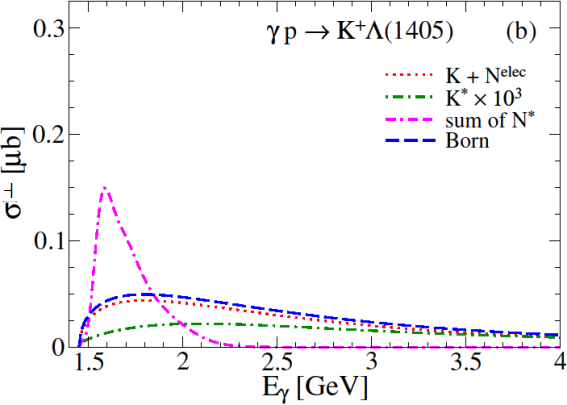}
\caption{(a) Parallel and (b) perpendicular cross sections for $\gamma p \to K^+
\Lambda(1405)$ are plotted as a function of the lab energy $E_\gamma$.
The curve notations are the same as Fig.~\ref{fig4}(a).
The $K^*$-Reggeon contribution is multiplied by the factor of $10^3$.}
\label{fig8}
\end{figure}
\begin{figure}[htp]
\centering
\includegraphics[width=7.0cm]{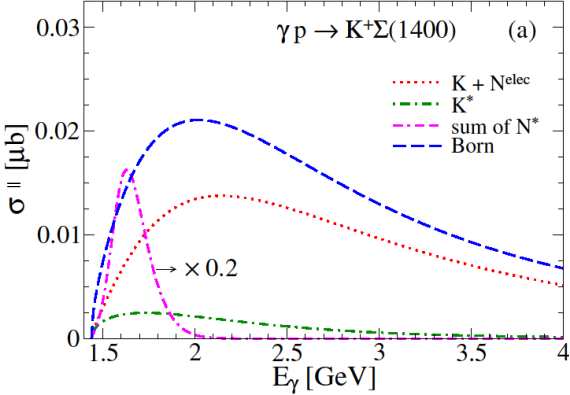} \hspace{0.3em}
\includegraphics[width=7.0cm]{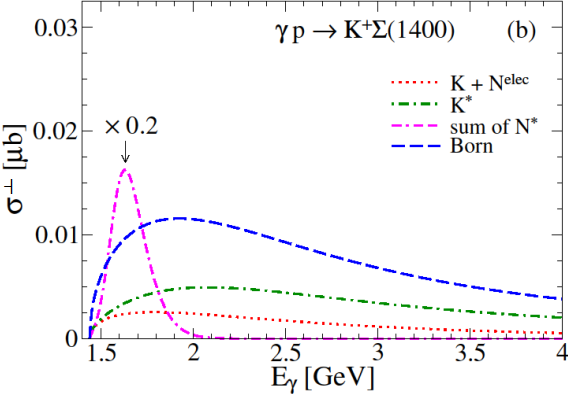}
\caption{(a) Parallel and (b) perpendicular cross sections for $\gamma p \to K^+
  \Sigma(1400)$ are plotted as a function of the lab energy $E_\gamma$.
The curve notations are the same as Fig.~\ref{fig4}(a).
The $N^*$ contribution is multiplied by the factor of 0.2.}
\label{fig9}
\end{figure}
Finally, we depict the energy dependence of the parallel ($\sigma^\parallel$) and
perpendicular ($\sigma^\perp$) cross sections in the left and right panels of
Figs.~\ref{fig8} and~\ref{fig9}, respectively.
As indicated from the discussion of the photon-beam asymmetry, $K$- and
$K^*$-Reggeon exchanges govern $\sigma^\parallel$ and $\sigma^\perp$, respectively.
It turns out the full result of $\sigma^\parallel$ for $\gamma p \to K^+ \Sigma^*$
is still an order of magnitude smaller than that for $\gamma p \to K^+
\Lambda^*$.
However, when it comes to $\sigma^\perp$, the full result for $K^+ \Sigma^*$
production is only 3 times suppressed when compared to that for $K^+ \Lambda^*$
production in the nonresonant region $E_\gamma \approx 3.0$ GeV.
This is because the $K^*$-Reggeon exchange for the former reaction is
approximately 100 times larger than that for the latter one.
Furthermore, the difference of $\sigma^\perp$ between two reactions becomes
smaller as $E_\gamma$ increases due to the larger intercept of the $K^*$
trajectory than that of $K$, i.e., $\alpha_{K^*}(0) > \alpha_K(0)$ (see
Eq.~(\ref{eq:ReggeTraj})).
Meanwhile, we  find that the $N^*$ contribution to $\sigma^\parallel$
and $\sigma^\perp$ is more or less equal to each other.

\section{Summary}
\label{SecIV}
In the present work we have investigated the reaction mechanism of the photoproduction of 
the  $\Sigma(1400)$ resonance found in Ref.~\cite{Khemchandani:2018amu}.
For a reasonable prediction of various physical observables we have 
studied the photoproduction of $\Lambda(1405)$ also, which has the same spin and
parity as $\Sigma(1400)$, i.e., $J^P = 1/2^-$.
We have employed an effective Lagrangian approach in the Born approximation, combining
it with a Regge model for the $t$-channel $K$- and $K^*$-Reggeon exchanges.
We have also considered contributions from the excitation of the nucleon in the intermediate state. 
The largest of such contributions comes from the inclusion of 
 $N^*(1895)$, to which two poles in the complex energy plane are associated;
$M_{N^*}-i\Gamma_{N^*}/2 = 1801-i96$ MeV and $1912-i54$ MeV.
The model parameters such as the cutoff mass and energy scale parameter are
the same for both reactions except for some coupling constants.

We have presented discussions on the couplings at strong vertices related to $N^*(1895)$ decay to
the $K\Lambda(1405)$ and $K\Sigma(1400)$ channels. 
We have also presented details on the determination of the radiative decay widths of $N^*(1895)$, $\Lambda(1405)$,
and $\Sigma(1400)$  by relying on the vector-meson dominance mechanism.

We have first obtained the total and differential cross sections for $\gamma p \to
K^+ \Lambda(1405)$ and compared them with the CLAS data.
Encouraged by the finding of a good agreement between our results and the data,
we have presented the predictions on the cross sections for $\gamma p \to K^+ \Sigma(1400)$.
The cross sections in the latter case have been found to be an order of magnitude suppressed 
because of the smaller value of $g_{KN\Sigma^*}$
relative to $g_{KN\Lambda^*}$.
Still, the magnitude of the  $\gamma p \to K^+ \Sigma(1400)$ cross sections found in our work, 
is large enough to be measured in the future experiments.
We have also found that the backward peaks in the differential cross sections for $K\Sigma^*$
production are relatively more visible than those for the $K\Lambda^*$ case due
to $\mu_{\Sigma^*\Lambda(\Sigma)} > \mu_{\Lambda^*\Lambda(\Sigma)}$.

Further, we  have shown the angle dependence of some polarization observables, such as
the photon-beam asymmetry, target asymmetry, and beam-target asymmetry.
We have found that the results for $K\Lambda^*$ production are clearly distinguishable from those
for the $K\Sigma^*$ case.
Such findings can play a crucial role in determining the corresponding reaction mechanisms.
A comparison of our results with  future experimental data can be useful in revealing a subtle
interference effect between the Born-term and resonance contributions.

We have finally presented the energy dependence of the parallel and perpendicular
cross sections and found that $K$- and $K^*$-Reggeon exchanges dominate
$\sigma^\parallel$ and $\sigma^\perp$, respectively.
The full result of $\sigma^\perp$ for $\gamma p \to K^+ \Sigma(1400)$ has been found to be only 3
times suppressed as compared to that for $\gamma p \to K^+ \Lambda(1405)$ at
$E_\gamma \approx 3.0$ GeV and the difference has been found to become smaller as $E_\gamma$
increases due to $\alpha_{K^*}(0) > \alpha_K(0)$.

\section*{ACKNOWLEDGMENTS}
This work is supported by the National Research Foundation of Korea funded by
the Ministry of Education, Science and Technology (MSIT) (NRF-2019R1C1C1005790
(S.H.K.), 2018R1A5A1025563, and 2019R1A2C1005697 (S.i.N.)).  K.P.K. and A.M.T. gratefully acknowledge the  support from the Funda\c c\~ao de Amparo \`a Pesquisa do Estado de S\~ao Paulo (FAPESP), processes No. 2019/17149-3 and 2019/16924-3, and from the Conselho Nacional de Desenvolvimento Cient\'ifico e Tecnol\'ogico (CNPq), grants No. 305526/2019-7 and 303945/2019-2. A.H. is supported in part by Grants No. 17K05441(C) and by Grants-in Aid for Scientific Research on Innovative Areas (No.\,18H05407).


\end{document}